\documentclass[sigconf]{acmart}

\usepackage{array}

\AtBeginDocument{%
  \providecommand\BibTeX{{%
    \normalfont B\kern-0.5em{\scshape i\kern-0.25em b}\kern-0.8em\TeX}}}

\copyrightyear{2021} 
\acmYear{2021} 
\setcopyright{acmlicensed}\acmConference[CHI '21]{CHI Conference on Human Factors in Computing Systems}{May 8--13, 2021}{Yokohama, Japan}
\acmBooktitle{CHI Conference on Human Factors in Computing Systems (CHI '21), May 8--13, 2021, Yokohama, Japan}
\acmPrice{15.00}
\acmDOI{10.1145/3411764.3445665}
\acmISBN{978-1-4503-8096-6/21/05}

\begin{document}

\title{Sharing Heartbeats: Motivations of Citizen Scientists in Times of Crises}


\author{Daniel Diethei}
\email{diethei@uni-bremen.de}
\affiliation{%
  \institution{University of Bremen}
  \city{Bremen}
  \country{Germany}
}

\author{Jasmin Niess}
\email{niessj@uni-bremen.de}
\affiliation{%
  \institution{University of Bremen}
  \city{Bremen}
  \country{Germany}
}

\author{Carolin Stellmacher}
\email{cstellma@uni-bremen.de}
\affiliation{%
  \institution{University of Bremen}
  \city{Bremen}
  \country{Germany}
}

\author{Evropi Stefanidi}
\email{evropi@uni-bremen.de}
\affiliation{%
  \institution{University of Bremen}
  \city{Bremen}
  \country{Germany}
}

\author{Johannes Schöning}
\email{schoening@uni-bremen.de}
\affiliation{%
  \institution{University of Bremen}
  \city{Bremen}
  \country{Germany}
}
\renewcommand{\shortauthors}{Diethei et al.}

\begin{abstract}
With the rise of COVID-19 cases globally, many countries released digital tools to mitigate the effects of the pandemic. In Germany the Robert Koch Institute (RKI) published the Corona-Data-Donation-App, a virtual citizen science (VCS) project, to establish an early warning system for the prediction of potential COVID-19 hotspots using data from wearable devices. While work on motivation for VCS projects in HCI often presents egoistic motives as prevailing, there is little research on such motives in crises situations. In this paper, we explore the socio-psychological processes and motivations to share personal data during a pandemic. Our findings indicate that collective motives dominated among app reviews (n=464) and in in-depth interviews (n=10). We contribute implications for future VCS tools in times of crises that highlight the importance of communication, transparency and responsibility.
\end{abstract}




\maketitle

\section{Introduction \& Motivation}

The ongoing COVID‑19 global pandemic~\cite{WHONovelCovid, huang_clinical_2020} has resulted in more than 29.3 million cases and 930,000 deaths as of September 15, 2020, in more than 188 countries~\cite{johns2020covid}. The virus is spread primarily via nose and mouth secretions~\cite{WHOCovidTransmitted}, and the two most common symptoms are fever and dry cough. However, symptoms can be non-specific, and some infected people can be asymptomatic~\cite{cebm_2020}. In conjunction with the virus' ranging incubation period~\cite{WHOQACorona}, these facts facilitate the (involuntary) transmission of the virus~\cite{cdc_coronavirus_2020}. 



Thus, governments, research facilities and industry have developed a variety of digital tools, which can fulfil central functions in the mitigation and control of the COVID-19 pandemic, by addressing different important aspects in the fight against it~\cite{whitelaw2020applications}. 
From a research point of view, technologies can be employed to control the pandemic, within five main application areas~\cite{vaishya2020artificial}: 
i) early detection and diagnosis of the infection,
ii) monitoring the treatment,
iii) contact tracing of the individuals,
iv) projection of cases and mortality, and
v) development of drugs and vaccines. 
Therefore, various mobile applications have emerged as digital tools aiming to accomplish the aforementioned goals~\cite{davalbhakta2020systematic}.

In particular, digital tools that are currently being utilised are covering the following areas:
(i) providing instant epidemic information (\emph{Johns Hopkins University Dashboard}~\cite{johns2020covid, dong2020interactive}; \emph{Worldometer's coronavirus update statistics}~\cite{worldometer2020}; WHO's \emph{Coronavirus Disease (COVID-19) Dashboard}~\cite{worldhealthorganizationdashboard}), 
(ii) providing health information and prevention advice (\emph{CovApp}~\cite{covapp2020}; India's \emph{GoK Direct}~\cite{GoKDirect}, Poland's \emph{ProteGO Safe}~\cite{ProteGOSafe}), 
(iii) facilitating contact tracing 
(Germany's \emph{Corona-Warn-App}~\cite{coronawarnapp}; Australia's \emph{COVIDSafe}~\cite{covidsafewikipedia_2020}; Singapore's \emph{TraceTogether}~\cite{tracetogetherwikipedia_2020}),
(iv) supporting symptom checking and feedback (Italy's \emph{Covid Community Alert}, Sri Lanka's \emph{Self Shield}~\cite{ListOfCoronaAppswikipedia_2020}), 
(v) monitoring quarantine compliance (Poland's \emph{Home Quarantine App}~\cite{hamilton_2020}; Karnataka's \emph{Quarantine Watch}~\cite{pti_2020}), 
(vi) or citizen science projects to support research (\emph{COVID Symptom Study research mobile app}~\cite{COVIDSymptomStudywikipedia_2020}, Germany's \emph{Corona-Data-Donation-App}~\cite{coronadatadonationapp}). 

The \emph{Corona-Datenspende-App} (CDA, \emph{Corona-Data-Donation-App}) \cite{coronadatadonationapp} is such a citizen science tool, published by the Robert Koch Institute (RKI; the German public health institute) in early April 2020 (see section~\ref{CDAsection} for a detailed description of the app and its functionalities). It provides citizens with the opportunity to share their health data from fitness trackers and smartwatches, with the aim to better record and understand the spread of COVID-19, detecting local fever outbreaks potentially associated with COVID-19~\cite{rkiDatenspendeApp}. This app is a virtual citizen science (VCS) project, meaning it comprises a crowdsourced approach conducted almost entirely through virtual portals~\cite{wiggins2011conservation}. We chose the CDA as our subject of research for multiple reasons. The CDA addresses a problem of global relevance, gathered a large user base (more than 500,000 users) within a few weeks of the release, was endorsed by the government and health authorities and was one of the earliest technological efforts in collecting physiological data for the detection of COVID-19 symptoms.

Understanding the motivations behind the participation in a VCS project that takes place under such exceptional circumstances is of great significance, especially given the current high stakes. The success of such research relies on the data collected, thus a higher number of people with continuing engagement are imperative for contributing and growing the data set. In the case of the CDA in particular, it is essential for volunteers to participate in the long-term.

In this paper, we study the motivations and investigate the socio-psychological processes which led people to engage in a VCS project and share personal data during the COVID-19 pandemic. In particular, we explore the "lived experience" of a VCS project and how it affected participation. The need to study the lived experience of such projects is also pointed out in Brown et al.~\cite{brown2015enjoying}. To that end, we researched the motivations of VCS users of the CDA.

We collected 10,202 reviews about the CDA from Apple's App Store and Google's Play Store (later described in detail in section~\ref{section:retrieval_interviews}) and coded 464 reviews to understand the motivations of the users, deriving a model about their lived experience and engagement lifecycle. Based on our findings, in a second step, we validated the model with in-depth interviews of ten app users, in which they discussed their experience using the CDA. Our findings indicate that, contrary to other VCS projects, collective motives dominated over egoistic ones. Moreover, we observed an unprecedented persistence from the app's users to continue using the CDA, despite various deterring factors (app instability and issues, negative media coverage etc). These motives and reasons are explored in-depth and presented in this paper, and we contribute design implications for future VCS tools in times of crises that embed user needs and aim for long-term engagement.

\section{Background \& Related Work}
In this section, we contextualise our research within past efforts. First, we review prior work about citizen science in HCI and Computer Supported Cooperative Work (CSCW). Then, we describe research on the motivational processes of social participation. 

\subsection{Citizen Science in HCI}
Citizen science, also known as community science, is scientific research conducted, in whole or in part, by nonprofessionals~\cite{gura_citizen_2013}. The definition of ‘citizen science’ and the terminology surrounding the concept is in constant flux \cite{eitzel2017citizen}. As outlined by Eitzel et al.~\cite{eitzel2017citizen}, citizen science encompasses a spectrum from active to passive participation. We adopt a definition in line with this broad vision. Citizen science can be described as public participation in scientific research and can lead to advancements in scientific research, as well as in an increase of the public's understanding of science~\cite{hand_people_2010, doyle_using_2019}. Citizen science can enable people to help solve problems that are relevant for or of interest to them~\cite{kim2011creek}, and most of its projects are contributory ones~\cite{rotman2012dynamic}. By participating as collaborators/volunteers in a scientific research project~\cite{irwin1995citizen}, citizen scientists are more engaged and have a more active role compared to scientific research studies involving members of the public as subjects. Moreover, by incorporating members of the public, citizen science projects enable data collection on a larger scale and over a wider geographical area~\cite{johnson2016not,johnson2017effect, rotman2012dynamic}. 
Citizen science can be employed in a variety of disciplines, such as biology~\cite{sullivan2009ebird}, environmental studies~\cite{kim2011creek}, and mathematics~\cite{cranshaw2011polymath}, with applications ranging from observation documentation~\cite{eBird} and data collection and analysis~\cite{zooniverse}, to content creation~\cite{nationalparksservice} and curation~\cite{encyclopediaOfLife}. 

In 2016, Preece~\cite{preece2016citizen} presented a research agenda on how HCI and citizen science can mutually enrich each other. Concurrently, within the last couple of years, HCI and CSCW scholars explored a variety of different aspects of virtual citizen science projects. Understanding how to facilitate and maintain engagement in virtual citizen science projects is one of the main questions HCI research focuses on. Eveleigh et al.~\cite{Eveleigh2014Dabblers} emphasised the need to study different groups of citizen science collaborators, namely highly committed volunteers, occasional contributors and contributors who drop out. They found that the type of motivation (i.e. extrinsic, intrinsic) was related to contribution behaviour. They determined that there is a need to design VCS interfaces that address a variety of different motivational needs of citizen scientists. 
Snyder~\cite{Snyder2017} explored visualisations created by and for nonprofessionals in the context of the citizen science project COASST (i.e. Washington’s Coastal Observation and Seabird Survey Team). Based on their results, they derived insights on how vernacular visualisations can enrich communication and coordination in citizen science projects. Moreover, scholar have explored the integration of gamification into VCS projects to increase user activity and support motivation. For instance, Eveleigh et al.~\cite{Eveleigh2013} found that competitive gamification elements led to various reactions from engagement to disengagement within the citizen science project. Based on their results, they derived design implications for citizen scientists with different engagement levels. Simperl et al.~\cite{SimperlGame2018} conducted a literature review combined with an interview study to identify the value of gamification elements for VCS projects. Their findings indicate that gamification elements appeared to have little impact on the recruitment of new citizen scientists. However, such elements seemed to be a means to maintain interest of already engaged citizen scientists. 
These examples indicate the need to build an understanding about the motivational processes of citizen science projects across various contexts (i.e. projects), encompassing citizen scientists with different levels of experience. Our work aims to take another angle from previous work as it looks specifically at the lived experience of a VCS project on the COVID-19 pandemic. Based on our results, we strive to derive implications to design VCS systems that embed user needs in times of personal and societal crises.

Another strain of research studied the social mechanisms of VCS projects. Jackson et al.~\cite{Jackson2020} described the dynamic process of volunteer engagement. Based on the example of the CS project Gravity Spy their study showed that citizen scientists initially engaged with resources by project organisers. However, their engagement shifted to community- and agent-centred resources in later stages of the project.
In contrast to most studies in VCS that focus on long-term engagement, Reeves and Simperl~\cite{Reeves2019} explored short-term citizen science projects. They noted that responding to discussion messages by volunteers was particularly important at the beginning of VCS projects to maintain involvement. On a similar note, Jay et al.~\cite{Jay2016} found that contribution rates to VCS projects can be significantly increased by allowing citizen scientists to contribute without registering for a designated citizen science project. Addressing the data privacy concerns in citizen science, Bowser et al. \cite{bowser2017accounting} report that values and norms of citizen science explicitly promote data sharing to achieve a greater good.
The examples outlined above all focus on topics that do not pose an immediate threat to the individual citizen scientist or the society as a whole (e.g. Old Weather project~\cite{Eveleigh2014Dabblers}, COASST~\cite{Snyder2017}). Our work aims to take another angle from previous work as it looks specifically at a virtual citizen science project that addresses an immediate crises. We strive to understand the lived experience of CDA as a VCS project to derive insights about the intricacies of designing VCS tools in times of crises.

\subsection{Motivations for Participation in Citizen Science}
Multiple studies have investigated which motives are responsible for participation in citizen science. Many report intrinsic motives, such as personal interest, as the most important factor for contribution~\cite{jennett2016motivations, nov2007motivates, rotman2014motivations}. However, some studies suggest that collective motives, such as contributing to science, motivate citizen scientists~\cite{raddick2013galaxy, curtis2015online}. Some studies found both intrinsic and collective motives make people participate in citizen science~\cite{nov2011dusting, land2016citizen}. 
Most studies on citizen science in HCI are based on motivational models from psychology, e.g. from Batson \& Ahmad~\cite{batson2002four}, who identified four types of motivations for social participation towards common goals: egoism, altruism, collectivism and principlism. \textit{Egoism} occurs when the ultimate goal is to increase one’s own welfare. \textit{Altruism} has the goal of increasing the welfare of another individual or group of individuals. \textit{Collectivism} has the goal of increasing the welfare of a specific group that one belongs to. \textit{Principlism} has the goal of upholding one or more principles dear to one’s heart (e.g. justice or equality). Batson \& Ahmad’s theory was not directed towards citizen science, but explains the mechanisms of motivation in building and sustaining a community~\cite{rotman2012dynamic}. 

Another model has been proposed by Klanderman, who explains voluntary participation in social movements~\cite{klandermans1997social}. This framework includes four classes of volunteers’ motivations for participation: collective motives (the importance attributed to the project’s objectives); norm-oriented motives (expectations regarding the reactions of important others, such as friends, family or colleagues); reward motives (benefits such as gaining reputation, or making new friends); and identification (identification with the group, and following its norms).
Klanderman's framework has been extended with an intrinsic motivation dimension, operationalised as the enjoyment associated with participation in the project. This extended framework has been applied in studies of open-source software development~\cite{hertel2003motivation} and Wikipedia editing~\cite{schroer2009voluntary}. Furthermore, reward motives have been divided into two specific motives: community reputation benefits and social interaction benefits~\cite{butler2002community, roberts2006understanding, nov2011dusting}. Rotman et al. \cite{rotman2012dynamic} describe a dynamic model of motivations that change during the contribution cycle of a citizen science project. Two important points in time are the initial encounter between a volunteer and scientific projects and the point at which a project ends. At the initial encounter, volunteers are driven by egoism. At the second end of a task or a project, other motivational factors, altruism and collectivism, become more important.

Citizen Science projects are conducted in different domains, and the motivations that drive participants vary. 

\subsection{Crisis Informatics}

A large array of past research efforts in HCI and beyond explored how people utilise technologies in crisis. Crisis informatics is an interdisciplinary research area investigating ‘the interconnectedness of people, organisations, information and technology during crises’~\cite{hagar2006using}. In recent years, crisis informatics increasingly focused on the sociotechnical intricacies connected to crises~\cite{Soden2018}. Previous work in crisis informatics inter alia explored topics such as public health crisis~\cite{Gui2018}, natural disasters~\cite{Chauhan2017,Vieweg2010} and climate change~\cite{Soden-et-al-2020}.

In the context of the Zika virus epidemic, Gui et al.~\cite{Gui2018} qualitatively studied the perception of risk and risk communication on Reddit. Their study provided relevant starting points for successful risk communication via social media. For instance, their results highlighted the relevance of personal experiences and the resulting perception of risks and risk measures. Besides, Gui et al. found that people also thought about the challenges facing society as a whole in times of crisis. Thus, they emphasised the need to foster dialogue between the authorities and citizens in times of crisis. The importance of another social media platform in times of crisis, namely Twitter, was explored by Vieweg et al.~\cite{Vieweg2010}. They analysed more than 20000 tweets connected to two natural hazards (the Red River Floods and the Oklahoma grass fires). Their work provided starting points on how social media could support affected areas in current crises. More precisely, Vieweg et al. discussed how information extracted from Twitter could increase situational awareness of the public and emergency responders. Similarly, Chauhan and Hughes~\cite{Chauhan2017} analysed online information on the Carlton Complex Wildfire. Their results emphasised the central role local media outlets played to distribute crisis information. They also discussed the increasing difficulty for those affected by a crisis to identify helpful information in the wealth of information available. 

On another note, Huang et al.~\cite{Huang2015} explored how the emotional proximity to an anthropogenic hazard, namely the Boston Marathon Bombings in 2013, affected information sharing behaviour. Their results showed that physical and emotional proximity to the crisis impacted how people sought and shared information online. Furthermore, the authors highlighted the critical role social media can play to those emotionally and physically affected by the crisis.

The aforementioned examples indicate a shift in crisis informatics from authority-centric ‘push’ culture towards citizens who are considered to be able to do life-saving work~\cite{tan2017mobile}. It has been suggested that future research should conduct user-centred studies for mobile crisis applications, specifically addressing the motivations of citizens. While there are usually multiple motivations involved~\cite{batson2002four, nov2011dusting}, the dominating factors differ between projects.

\begin{figure*}[t]
  \centering
  \includegraphics[width=\textwidth]{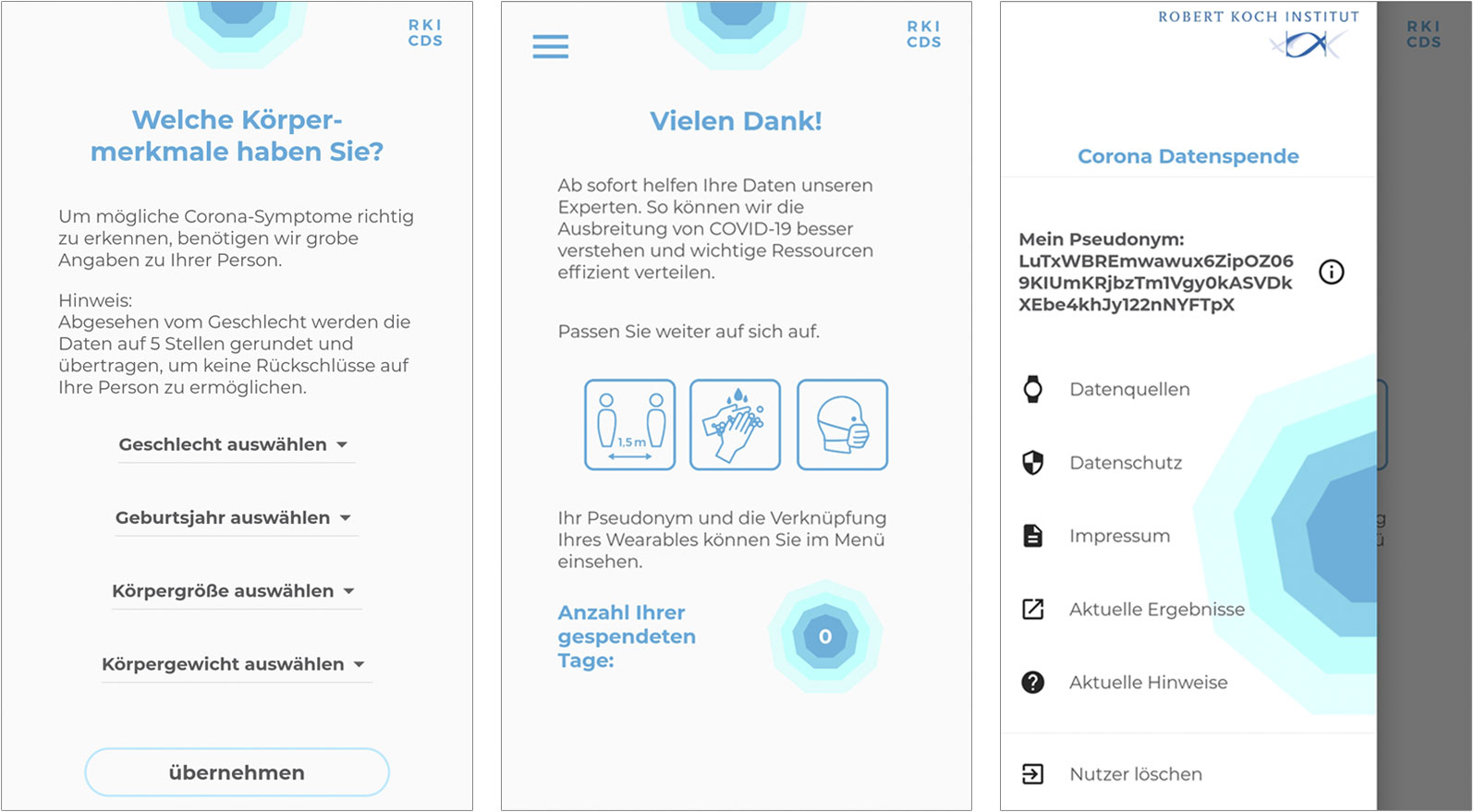}
  \caption{Three screenshots of the CDA on a smartphone in German, the only available language for the app. The following descriptions are from top to bottom in each screenshot. Left: One of five screens of the app's initial setup asking users for their gender, birth year, height and weight. Middle: Homescreen of the CDA thanking users and showing a short explanation of the app's purpose, three graphics explaining three important measures to mitigate the transmission of the virus (1.5m physical distance, washing hands, wearing a mask), a hint that the personal pseudonym and the wearable's registration setting can be found in the menu and the current number of donated days, here "0". Right: Menu of the CDA displaying the user's pseudonym and offering access to the wearables registration settings, the apps' data privacy and imprint, current information on the results (leads to the blog on the CDA website), current information about the app and an option to delete the user account. The remaining screens of the CDA, which can be accessed through the menu, do not offer (other) interactive elements or information regarding the user's data donation.}

  \Description{Three screenshots of the CDA on a smartphone in German, the only available language for the app. The following descriptions go from top to bottom in each screenshot: (left) One of five screens of the app's initial setup asking users for their gender, birth year, height and weight. (middle) Homescreen of the CDA thanking users and showing two sentences about the app's purpose, three graphics explaining three important measures to mitigate the transmission of the virus (1.5m physical distance, washing hands, wearing a mask), a hint that the personal pseudonym and the wearable's registration setting can be found in the menu and the current number of donated days. (right) Menu of the CDA displaying the user's pseudonym and offering access to the wearables registration settings, the apps' data privacy and imprint, current information on the results (leads to the blog on the CDA website), current information about the app and an option to delete the user account. The remaining screens of the CDA, which can be accessed through the menu, do not offer (other) interactive elements or information regarding the user's data donation.}
  \label{fig:app_screenshots}
\end{figure*}

Our work is also inspired by previous work in HCI that focuses on long-term user experience (UX). A crisis itself or the consequences of a crisis are often long-lasting. In line with that, ideally, the CDA should be used over a more extended period of time. As our study focuses on the lived experience of the CDA, there is a need to discuss previous studies on long-term UX. Karapanos et al.~\cite{Karapanos2009} introduced a conceptual model that focused on the temporality of user experience. Their model encompasses three forces, familiarity, functional dependency and emotional attachment. The three forces impact how people transition across three different phases of their experiences with an interactive product. The phases are orientation, incorporation and identification. Since we focus on the lived experience of the CDA, the third phase identification and the associated force emotional attachment is of particular relevance in our inquiry. The identification phase focuses on how an interactive product is integrated into ones’ everyday life and concurrent emotional experiences. Consequently, we consider emotions as a key variable in our study. In the field of personal informatics, Epstein et al.~\cite{Epstein2015}, introduced the lived informatics model of personal informatics. Also, Epstein et al.~\cite{Epstein2016} emphasised the importance of emotions during and after the personal informatics experience, such as feeling guilty for stopping to track. Even though using the CDA in a pandemic is not the same as a lived informatics experience described by Epstein et al.~\cite{Epstein2015}, there are some similarities, such as collecting and sharing personal data. Thus, in our study, we also focus on the entire lived experience of the CDA. To summarise, while citizen science projects in crises have been described in the crisis informatics literature, to the best of our knowledge, no prior work in HCI addressed emotions and motivations surrounding citizen science projects in times of crises. Hence, our work explores the lived experience of the CDA with a focus on emotions and motivations.


\section{Research Questions}
Our overarching goal is to learn more about the lived experience of the Corona-Data-Donation App. We endeavour to understand the motivations of people contributing to citizen science projects that focus on incidents posing an immediate threat to the individual citizen scientist or the society as a whole. We explore this topic by focusing on the CDA, a virtual citizen science project that aims to mitigate the impact of the COVID-19 pandemic. To that end, we strive to answer the following research questions:  
\begin{itemize}
    \item RQ1: What are the motivations to participate in a VCS project in times of crises?
    \item RQ2: What are the socio-psychological processes associated with the participation in a VCS project in times of crises?
\end{itemize}



\section{Method}\label{method_section}
This study explores the lived experience of the CDA. To that end, we applied a two-step procedure. We retrieved and qualitatively analysed users' online reviews of the CDA from the App Store and the Play Store to derive a holistic understanding about the experience with the CDA. As a second step, we conducted semi-structured interviews with ten individuals to explore themes identified in the reviews in greater detail. The reviews helped us to identify the range of different motivations present among a large group of users, however, we anticipated the reviews to be of a rather strong opinion. The additional analysis of the interviews allowed us to ask differentiated questions on specific motivations that contributed to a more nuanced picture of the users' reasoning for participation. Additionally, as one of our research goals targets the socio-psychological processes of citizen scientists, the complementary analysis gave us two perspectives on social interaction among users. First, reviewers can interact with other reviewers and the developers in the app stores. Second, we asked our interview partners about the exchanges they had with family and friends about the CDA. Our understanding of the lived experience of the CDA is based on a qualitative analysis of these two data collections.

\subsection{Corona-Datenspende-App}\label{CDAsection}

The CDA~\cite{coronadatadonationapp} (Corona-Datenspende-App) app was released on April 7th 2020 by the RKI which has worked together with the German government to coordinate Germany’s response against COVID-19. The general aim of the CDA app is to monitor the spread of COVID-19 and analyse whether measures to contain the novel coronavirus pandemic are working~\cite{busvine_2020}. The app gathers vital signs from volunteers wearing smartwatches or fitness trackers - including heart rate, daily steps and sleep habits - to analyse whether they are symptomatic of the flu-like illness. They can indicate fever, e.g. when the resting heart rate goes up and the daily steps go down. However, these physiological attributes do not necessarily mean that the user has developed a fever; at the same time, multiple causes for a fever exist, so the data has to be supplemented by other sources, for example COVID-19 case numbers. So far, there is limited evidence that passive physiological data alone support the detection of COVID-19~\cite{quer2020passive}. Preliminary results of the CDA are reported in a blog and in an interactive fever curve on the CDA website with the aim to help the health authorities and the general public to assess the prevalence of infections down to state level, in combination with other data inputs~\cite{feverCurve}.

As of September 15 2020, 526,727 people downloaded the app and registered a wearable device~\cite{rkiThankYou}. During the CDA's initial setup process, users are asked to provide their information on postal code, gender, birth year, height and weight (see Figure~\ref{fig:app_screenshots} (left)) and to register their wearable. After completing these initial steps, the physiological data measured by the wearable is automatically transmitted to the RKI from the user account at the wearable's provider~\cite{CCCdatenspende}. Thus, the data donation begins (see Figure~\ref{fig:app_screenshots} (middle)) and no user interaction with the app is required after this. The app's menu, however, offers access to further information on the user's personal pseudonym, data privacy, imprint, current information related to the research results (leads to the blog on the CDA website) and current information about the app (see Figure~\ref{fig:app_screenshots} (right)). Users can also change or disconnect their wearable or delete their account.

\begin{figure*}
  \centering
  \includegraphics[width=.93\textwidth]{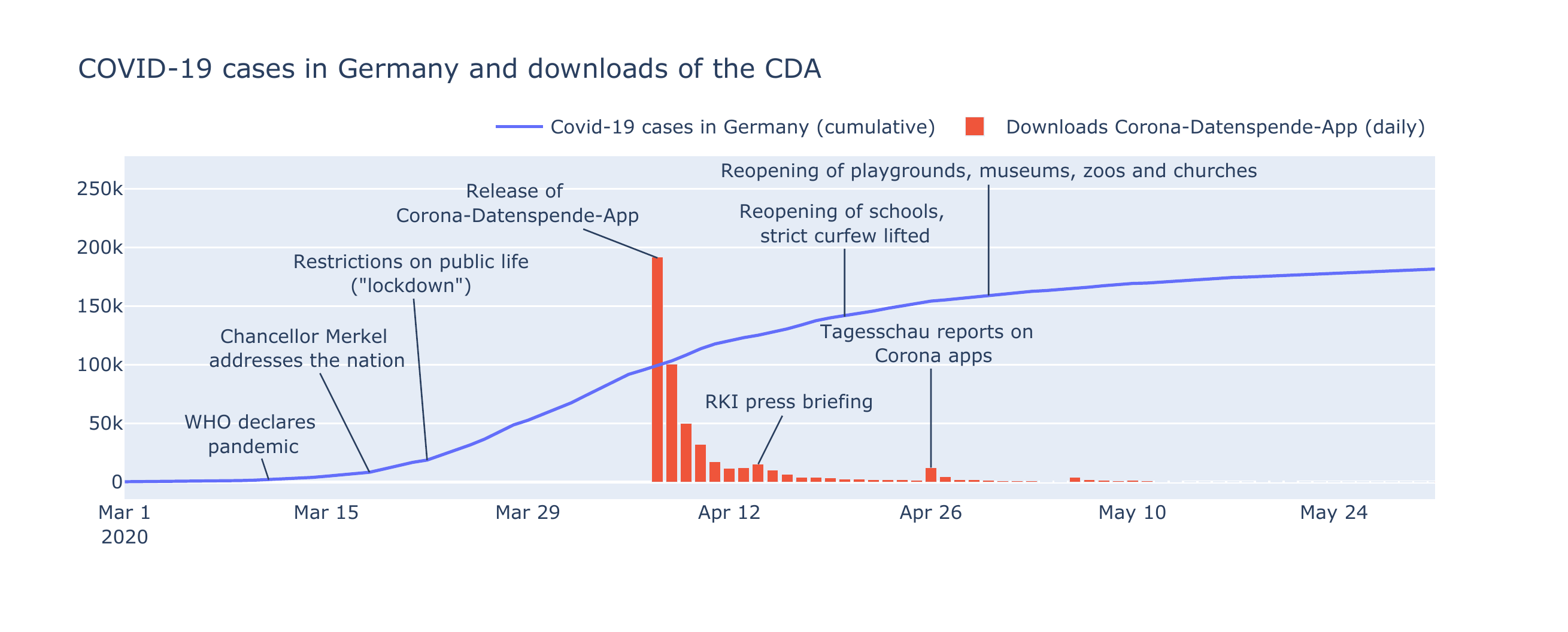}
  \caption{COVID-19 cases in Germany and downloads of the CDA. The CDA was released early in the pandemic, when cases were rising rapidly. Two small spikes can be seen after the RKI gave a press briefing and when the German public broadcasting news show \textit{Tagesschau} reported about German Corona Apps and showed an image of the CDA on national television. On that day, Tagesschau has actually only reported about Corona-Warn-App but showed an image of the CDA with the name clearly visible. Sources: mdr.de \cite{MDR2020chronik} and rki.de \cite{rkiThankYou}.}
  \Description{COVID-19 cases in Germany and downloads of the CDA. The CDA was released early in the pandemic, when cases were rising rapidly. Two small spikes can be seen after the RKI gave a press briefing and when the German public broadcasting news show \textit{Tagesschau} reported about German Corona Apps and showed an image of the CDA on national television. On that day, Tagesschau has actually only reported about Corona-Warn-App but showed an image of the CDA with the name clearly visible. Sources: mdr.de \cite{MDR2020chronik} and rki.de \cite{rkiThankYou}.}
  \label{corona_in_germany}
\end{figure*}

The CDA was published by the RKI which is a federal institute under the authority of the Federal Ministry of Health of Germany. The call to use the app was backed up by the German government~\cite{bundesregierung2020datenspende} 
and multiple news channels reported on the day of release about the CDA on national television~\cite{tagesschau2020datenspendeTV, ZDF2020datenspendeTV}, their online platforms~\cite{tagesschau2020datenspendeOnline, ZDF2020datenspendeOnline} and continued to report, e.g. about the reached user number of half a million after one month~\cite{tagesschau2020datenspende}. Two weeks after the app's release the Chaos Computer Club (CCC) analysed the CDA and publicly criticised parts of its data privacy practice~\cite{CCCdatenspende}. For example, the fitness data was only pseudonymised after being uploaded to the RKI servers. As a response, the RKI started to immediately address these issues in the implementation to close the related safety gaps.
Media coverage had a visible impact on download rates as well as reviews provided (see Figure \ref{corona_in_germany}). This paper focuses on the CDA, which was released before and should not be confused with the Corona-Warn-App~\cite{coronawarnapp}, a contact-tracing application launched by the RKI, utilising Bluetooth and location tracking.



\subsection{Data Collection \& Analysis - CDA App Reviews}
\label{section:retrieval_cda}
Our original data set consisted of 10,202 reviews from the App Store and Play Store. The reviews were submitted in the period between April 7 (April 9 for the App Store), when the apps were published, and April 26, which was the last day with more than 100 reviews on both platforms. 
For the Play Store, we scraped reviews with the python package \textit{google\_play\_scraper}. For the App Store we accessed the RSS feed which only contained the most recent 500 reviews.
The reviews information included a unique review ID, submission date, title (App Store only), review text, vote sum, vote count, score, version, author, reply content and reply date. All reviews were imported into the MAXQDA data analysis software. 

We applied open coding combined with thematic analysis similar to Blandford et al.~\cite{blandford_qualitative_2016}. Previous work in the field of HCI has utilised app reviews in multiple ways, including to characterise attitudes of users and functionality of apps~\cite{Stawarz2014pill, ghosh2018parental}. Extracting user experience information from online reviews has been, inter alia, done for software and video games~\cite{hedegaard2013extracting}. Similarly, in this work, we retrieve the experience of users of the CDA based on app reviews.
According to Vaast~\cite{vaast2017building}, it is important for researchers to access the context of the texts they are analysing. Some online texts, e.g. tweets, are very short (140 characters or less), thus making sense of them in and of itself is difficult. Therefore, tweets are to be understood within an ensemble~\cite{boyd2010tweet}, e.g. within surrounding tweets. 

Accordingly, we analysed the data in the context of the pandemic, meaning that we were aware that the motivations of the users differed from previous encounters with similar apps that collected physiological data for no direct personal benefit. 

In total, we coded 464 reviews from our dataset of 10,202 reviews. Three coders coded the data independently. We selected and coded the 464 reviews in four iterations, resulting in four samples, and updated our coding tree after each iteration until theoretical saturation was reached \cite{blandford_qualitative_2016}. Sample \#1 consisted of 84 reviews, which were randomly sampled from our dataset. This was followed by iterative discussions to establish the initial coding tree. For samples \#2 (108 reviews), \#3 (136 reviews) and \#4 (136 reviews) we followed the same procedure: For each app store (App Store and Play Store), we selected seven versions of the CDA, all ratings (5) and sampled two reviews for each rating. This gives us a maximum of 140 reviews for each sample. Since not all versions had at least two reviews for each rating, e.g. version 1.0.7 in the App Store had only 22 reviews, none of which were 2 or 4-star ratings, the actual amount of reviews differs (e.g. just 108 in sample 2). Some versions had less than two reviews for a rating, e.g. version 1.0.7 in the App Store had only 22 reviews, none of which were 2 or 4-star ratings. Therefore, the actual sample sizes vary and are below 140. 


\subsection{Data Collection \& Analysis - Interviews CDA Users}
\label{section:retrieval_interviews}

\begin{table}[b]
\centering
\begin{tabular}{ m{1.5cm} | m{1cm} | m{1cm} | m{3cm}}
\textbf{Participant} & \textbf{Age} & \textbf{Gender} & \textbf{Profession}\\
\hline
P1 & 27 & female & PhD student\\
P2 & 22 & female & student (medicine)\\
P3 & 22 & female & student (psychology)\\
P4 & 29 & male & PhD student\\
P5 & 23 & female & student\\
P6 & 39 & male & PhD student\\
P7 & 24 & female & student (psychology)\\
P8 & 23 & female & student\\
P9 & 20 & male & student\\
P10 & 35 & male & engineer\\
\end{tabular}
\caption{Demographics of interview participants.}~\label{tab:demo}
\end{table}
\begin{figure*}[t]
  \centering
  \includegraphics[width=\textwidth]{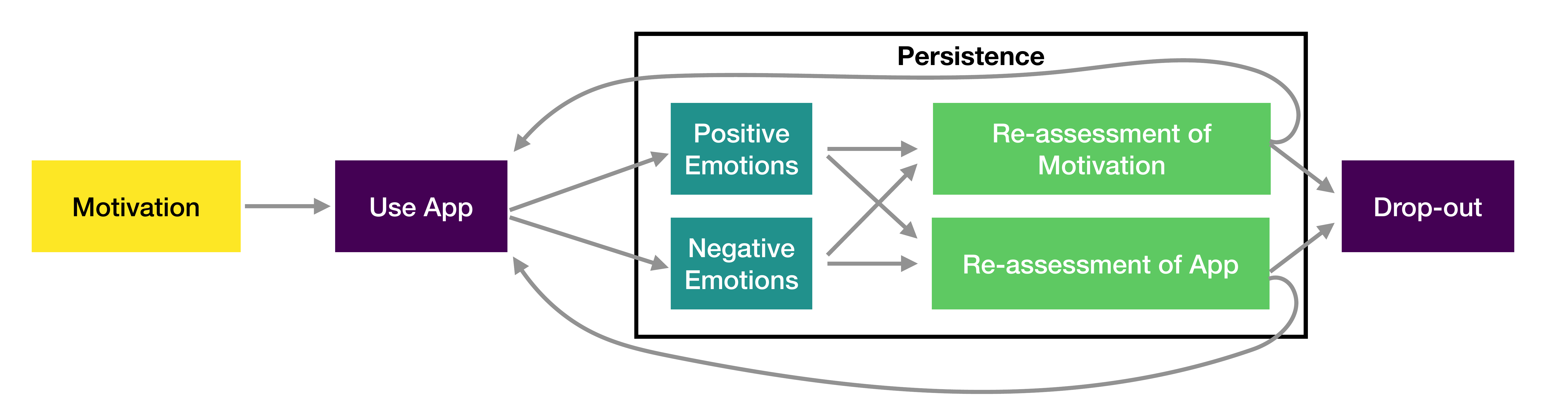}
  \caption{The experience of using the CDA, a virtual citizen science project, during the COVID-19 pandemic. People's entry point is the initial \textit{motivation} leading to the decision of \text{using the app}, i.e., participating in the VCS project. After the download and installation, users are confronted with the setup phase and exposed to positive and negative factors of the app. Consequently, \textit{positive and negative emotions} arise causing a \textit{re-assessment of the motivation} and \textit{re-assessment of the app}. Based on users' decisions, what follows is either a drop-out, i.e. removal of the app or cancellation of the data donation, or continued usage of the app. Particularly in the case of experienced negative emotions and the decision to continue donating data, users demonstrated persistence in remaining a citizen scientist.}
  \Description{The experience of using the CDA as a citizen scientist during the COVID-19 pandemic. People's entry point is the initial \textit{motivation} leading to the decision of \text{using the app}, i.e., participating in the VCS project. After the download and installation, users are confronted with the setup phase and exposed to positive and negative factors of the app. Consequently, \textit{positive and negative emotions} arise causing a \textit{re-assessment of the motivation} and \textit{re-assessment of the app}. Based on users' decisions, what follows is either a drop-out, i.e. removal of the app or cancellation of the data donation, or continued usage of the app. Particularly in the case of experienced negative emotions and the decision to continue donating data, users demonstrated persistence in remaining a citizen scientist.}
  \label{fig:flowchart}
\end{figure*}
To complement our review data, we recruited 10 users of the CDA app through the \textit{Prolific} platform and various Facebook groups. 
Participants' age range was 20-39 (\textit{M} = 26.4, \textit{SD} = 6.22), with six female and four male participants. The majority of the participants (8/10) were students, including three PhD students, three psychology students and one medical student. Other occupations included research assistant and engineer. The participants were compensated with 8€, either through an Amazon voucher or directly through \textit{Prolific}. Table \ref{tab:demo} presents details about the participants. 

All interviews were conducted via Zoom or via telephone using audio-only recording. Participants were asked for consent for recording before the interviews. The interviews lasted an average of 15 minutes (min = 08:40, max = 22:05, sum = 02:15:38). Using a semi-structured approach, we covered the topics of motivation, experience (e.g. installation issues and frequency of use) and data (e.g. data privacy, awareness of types of data shared). Most questions were open-ended, e.g. ``please describe your experience from downloading the app until today''. We provide the interview guideline as a supplementary material. All interview recordings were transcribed verbatim and imported into the MAXQDA data analysis software.

\section{Results}
In this section, we report on our findings from the analysis of the reviews and the interviews. First, we present the descriptive analysis of the reviews. Second, we explain the experience of the CDA citizen scientists and the three themes we identified in the reviews and interviews, i.e. motivations, emotions and persistence. We provide the coding tree in the supplementary materials. 

The experience of citizen scientists using the CDA during the COVID-19 pandemic is shown in Figure~\ref{fig:flowchart}. Individual aspects of this cycle of engagement demonstrated with the CDA such as the motivation and persistence are highlighted in the following subsections. Direct quotes were translated from German.

\subsection{Reviews}
\begin{figure*}[t]
  \centering
  \includegraphics[width=\textwidth]{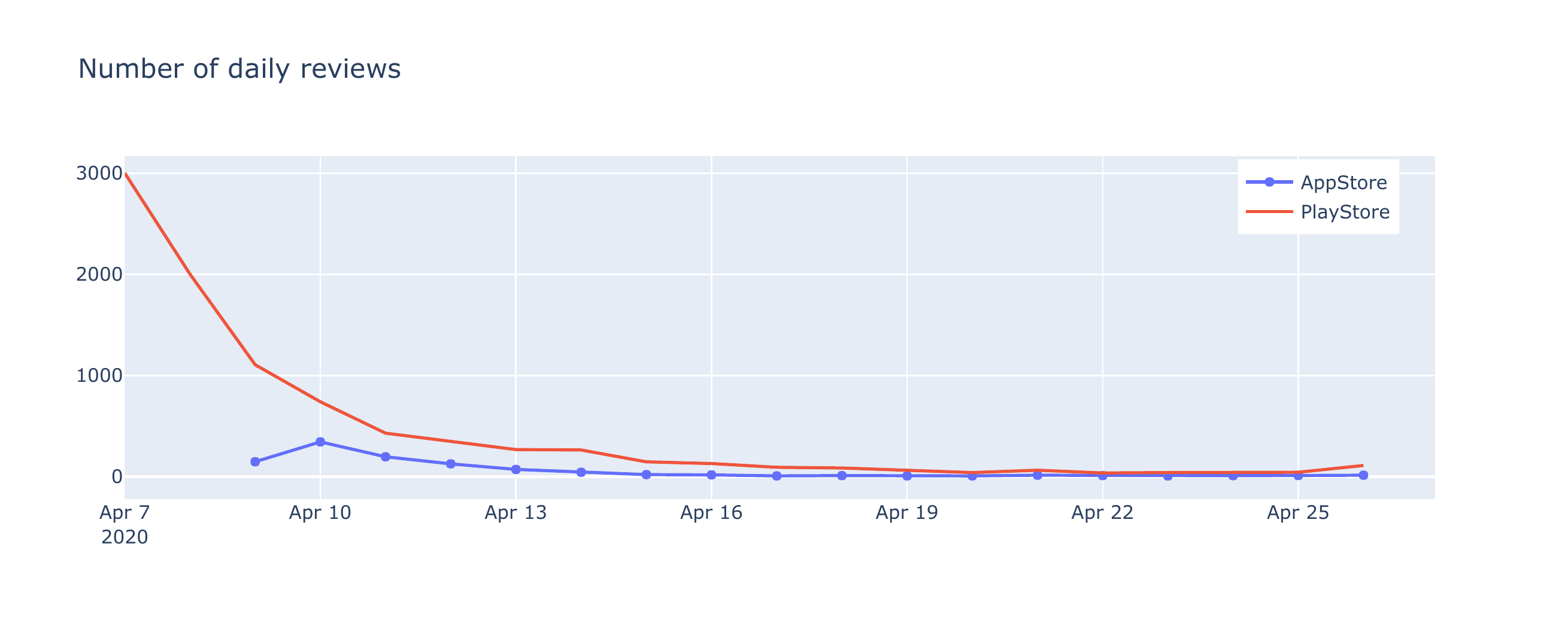}
  \caption{Number of daily reviews submitted for the CDA between April 7 and April 26 (Google Play Store), and April 9 and April 26 (Apple App Store). More than 80\% of the reviews were submitted in the first seven days after publication of the CDA.}
  \Description{New daily reviews submitted for the Corona-Data-Donation-App.}
  \label{new_daily_reviews}
\end{figure*}

Our analysis included 464 reviews in total in equal proportions from the App Store and Play Store. The average number of words per review was 41.8 (\textit{SD} = 29.7) in the App Store (title and review combined) and 22.0 (\textit{SD} = 17.4) in the Play Store (review only). Most of the reviews were submitted within the first days after publication. For example, the number of daily published reviews in the Play Store peaked around 3,000 on the first day -- almost a third of all collected reviews (Figure \ref{new_daily_reviews}). Afterwards, the submission rate dropped rapidly.

\begin{figure*}[t]
  \centering
  \includegraphics[width=\textwidth]{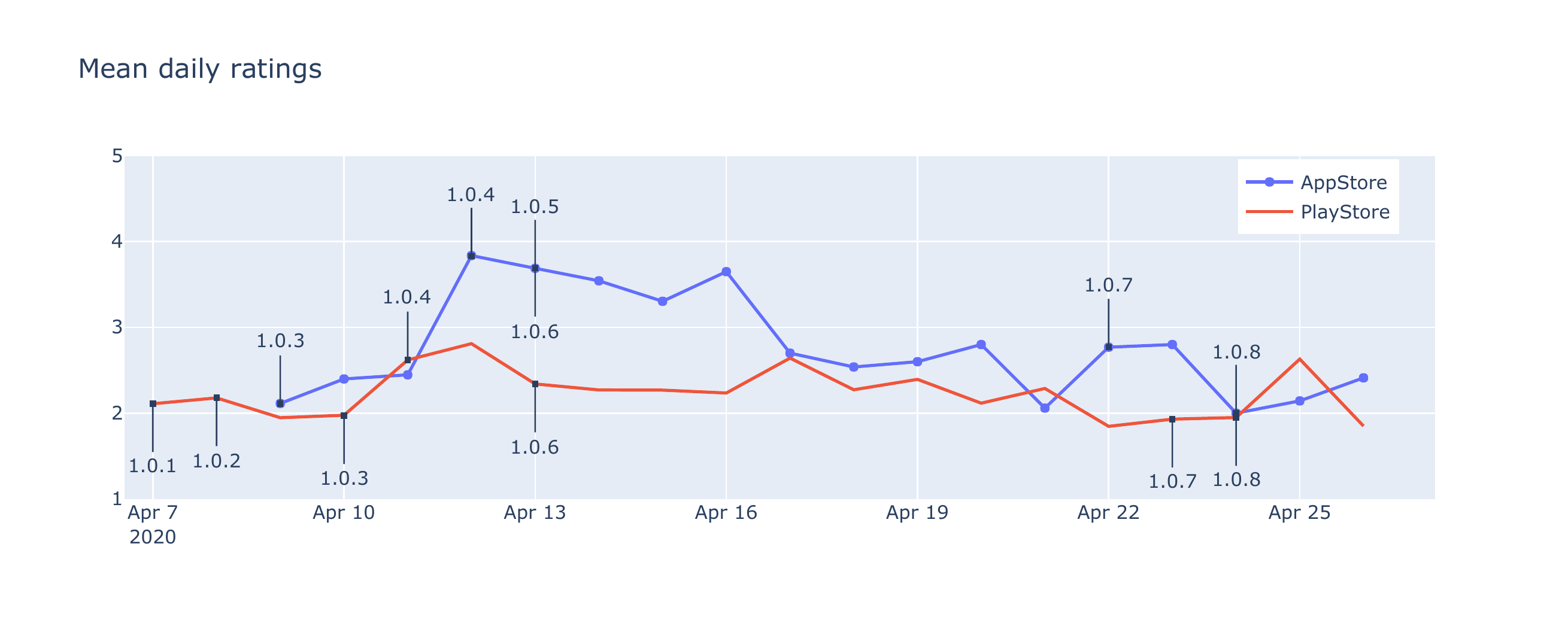}
  \caption{Mean daily ratings of reviews submitted for the CDA between April 7 and April 26 (Play Store), and April 9 and April 26 (App Store). Annotated are the version numbers. Version 1.0.4 fixed a common error with the day counter as can be seen by the increase in the scores. The App Store ratings were generally slightly higher compared to the Play Store.}
  \Description{Mean daily ratings of reviews submitted for the CDA between April 7 and April 26 (Play Store), and April 9 and April 26 (App Store). Annotated are the version numbers. Version 1.0.4 fixed a common error with the day counter as can be seen by the increase in the scores. The App Store ratings were generally slightly higher compared to the Play Store.}
  \label{mean_daily_ratings}
\end{figure*}

\begin{figure*}[t]
  \centering
  \includegraphics[width=\textwidth]{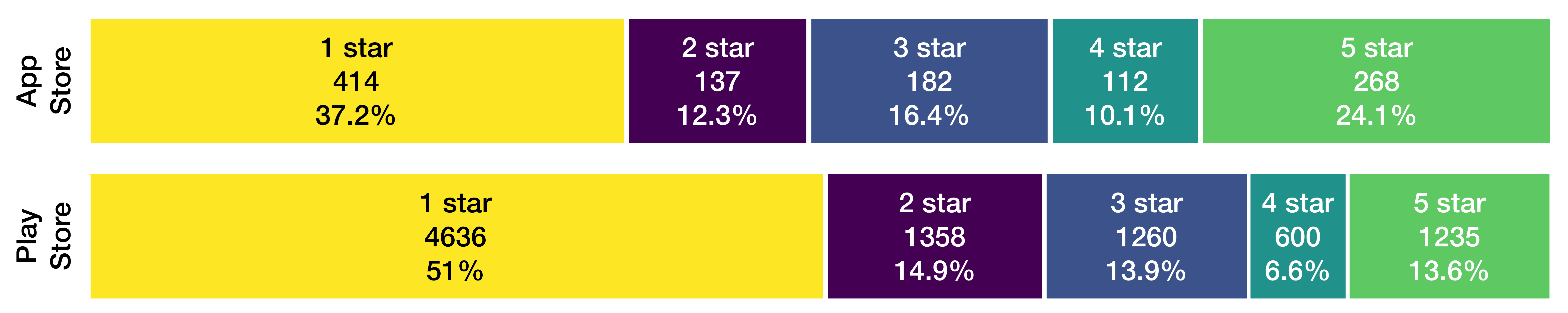}
  \caption{Distribution of ratings for both stores. Half of the Play Store ratings were 1-star, whereas only a third of the App Store reviews were 1-star.}
  \Description{Main daily rating of reviews submitted for the Corona-Data-Donation-App.}
  \label{proportion_ratings}
\end{figure*}

The mean ratings, illustrating users' quantified feedback between 1 and 5, were generally slightly higher in the App Store compared to the Play Store (Figures \ref{mean_daily_ratings}). Reviews were mostly rather negative or rather positive (Figure \ref{proportion_ratings}). Ratings went up after the release of the version 1.0.4 on April 12 which fixed an error regarding a broken counter for the number of donated days. A slight downward trend followed afterwards.

\subsection{Initial Motivation}
Initial motivations for participation were collective and mainly revolved around the opportunity to serve the community in an unprecedented situation and the role of the RKI as a trusted organisation.
These collective motivations were expressed through a variety of perspectives, such as an explicit desire to support the project due to the pandemic and its societal relevance. Many users expressed that they were "happy to support" the research of the RKI to better understand COVID-19 in such an unprecedented difficult situation. One common view amongst users was that the app enabled individuals to contribute to the fight of stopping the spread of the COVID-19 pandemic. The motivation of some users to serve their community overcame their concerns about sharing privacy-sensitive data. While the following quote also includes the potential personal benefits of donating data, i.e. an egoistic motive, it highlights the main motivation is still to serve the community.
\begin{quote} 
\emph{I had never thought I would share such intimate details, but everything that now helps statisticians to understand [the shared data], protects my family members at risk and friends. Eventually, also me! (App Store, 1.0.4, 5 stars)}
\end{quote}

Another factor was the hope for the intended success of the app which could lead to lifting social distancing measures in the long run. This was valued higher by one user than the data privacy, which was considered to be a "luxury product" in this unfamiliar situation. Other reasons for the willingness to share personal data were the pseudonymisation of the data, the trust in the RKI and general concerns about the community's health. As mentioned frequently in the interviews, even users who did not usually share their physiological data with other services participated in the donation (P2, P7, P8, P9), emphasising the trust in the RKI, both in terms of data privacy and hopes for mitigating the pandemic:
\begin{quote} 
\emph{In contrast to giving the data to Google, they go to the government. And I think they handle the data significantly better compared to when the data are sold. And this I considered a small opportunity to help. (P9)}
\end{quote}

The commitment to serve for the ``greater good`` was also accompanied by requests to share more data than the app asked for, e.g. the fitness data collected by their devices before the app was downloaded. In addition, it was requested to manually insert information about temperature, pregnancy status, allergies, general health status or symptoms. P8 expresses readiness to share sensitive data: 
\begin{quote} 
\emph{Yes, whatever they had asked for. Everything. In my head there are symptoms, like how I feel, if I have a high temperature. Maybe also normal things, such as how often do you go outside, so you can maybe detect preventively and protect [yourself] from Corona, not only [detecting] symptoms.'' (P8)}
\end{quote}
One user even reported purchasing a wearable for the sole purpose of donating data to the CDA.

Often, users projected their motivation to support the project to other users. This became apparent, when they peer-pressured others in their reviews to help the RKI and accept certain errors or to also donate their data. 
\begin{quote} 
\emph{Let's be reasonable and donate data! (App Store, version 1.0.6, 5 stars)}
\end{quote}
\begin{quote} 
\emph{Help the RKI and participate! Accept bugs in the short-term. (Play Store, version 1.0.6, 5 stars)}
\end{quote}

Responses from the interviews provided a deeper insight into certain motivational factors. Consistent with findings from the reviews, interview participants indicated that collective motives dominated the decision to contribute. Participants were aware that they had no direct benefit from using the app, but could potentially support the community. While the pandemic was considered serious by all interviewees, concerns mainly focused on other community or family members who were at risk. Two interview participants, P1 and P9, downloaded the CDA while searching the Corona-Warn-App since it appeared in the search results and was also launched by RKI. This indicates that their initial motivation might have been egoistic or collective, since the Corona-Warn-App warns of contact with infected people, which benefits both the individual and society. In this case it is difficult to distinguish between egoistic and collective motives.

Overall, a typical behaviour observed in the reviews consisted of a general appreciation of the app's purpose and, consequently, a strong initial motivation to contribute. 
\begin{quote} 
\emph{The idea is great, but unfortunately no way to contribute with my Samsung Gear Sport. (Play Store, version 1.0.1, 3 stars)}
\end{quote}
However, this was often followed by frustration due to errors or the incompatibility of many devices excluding users from donating their data.

\subsection{Emotional Responses to Problems}
Having charted the motivations of citizen scientists at the start of their contribution, we now present the emotional responses when being exposed to errors in the app. The experience of sharing data through the CDA was accompanied by positive and negative emotions. A common perception was the wish to help and the consequent disappointment when facing errors that prevented the contribution, mainly caused by a lack of feedback in the app and a limited range of supported devices.
A recurrent theme in the reviews was a positive attitude towards supporting the fight against COVID-19. A few users expressed their gratitude for the "hard work to manage this crisis". Users were also thankful for updates that resolved issues with the app. We will explain the positive emotions in more detail in the next section.

It became obvious in the reviews that the app was released with some errors that prevented or complicated the intended use for a large proportion of the reviewers. This circumstance caused different negative emotions such as frustration, anger or sadness.
In particular, many reviews addressed the displayed counter of donated days that was stuck at zero days. Users were, therefore, uncertain and sceptical if their data donation was successful, i.e. if they were actually helping or if an error in the app's development hindered their wish to contribute. This unawareness and lack of feedback about the state of the data donation caused frustration, and even when the day counter became active, some users were not entirely convinced about the correct functioning of the system:
\begin{quote} 
\emph{After the update, I can now finally see the days. Thanks a lot. However, information about which data are sent and what their purpose is is still missing. (App Store, version 1.0.4, 3 stars)}
\end{quote}

Another dominant point of criticism leading to frustration was the limited range of supported fitness trackers and smartwatches. Many users were unable to contribute due to their own unsupported device and communicated  their frustration about this issue understanding, especially after a month of waiting for an update. In some cases, frustrated users reported that this only became clear to them at the end of the registration process. Users were also frustrated that the CDA was only usable in combination with a fitness tracker or smartwatch, excluding many people who were not able to afford these devices. 

While technical difficulties were the predominant reason for negative emotional responses towards the app, we also observed concerns about transparency. In particular, most users were left unaware about the exact kind of data being collected and how it would be used, as the RKI reported the generic types of data the app requested, e.g. daily steps and heart rate, on the CDA website but not in the app. Additionally, there was no option to see what data an individual had shared with the RKI.

Overall, users often had hoped to contribute, but were excluded due to reasons out of their control. A common expression in the reviews was "I would like to help, but" which was followed by the explained barriers. This led to sadness, frustration and anger, expressed in disappointment and threats to stop the data donation. 

In response to the experienced errors, some users also expressed anger about a lack of professionalism in the app development and, based on that, doubted that their data was handled securely enough. Other negative comments related to the app drawing significant battery or appeared in the context of misconceptions about the CDA.
For example, one user was frustrated that the CDA would not alert users in case an infection was suspected based on the data they shared.
\begin{quote} 
\emph{But, what annoys me enormously are the explanations on the RKI website\footnote{The information this user refers to is found on the CDA website, which is provided by the RKI, but not on the RKI website.}: "The Corona data donation app is not a Corona test. The users themselves are not informed about a possible infection." No one in their right mind will assume that a corona test is possible with the data from a "fitness tracker". It would be nice if the app would inform users about "anomalies" (heart rate too high in conjunction with short deep sleep phases). Theoretically, I can deduce that myself, but this way the app would provide a benefit for the user. (App Store, version 1.0.4, 2 stars).}
\end{quote}


\subsection{Persistence}
Our analysis showed that, after using the app, users re-assessed their motivations and attitudes towards the app (see Figure \ref{fig:flowchart}). Initially, the shortcomings of the app led to frustration and drop-outs for many users. However, when contributors reassessed their motivations, they reminded themselves of their initial motivation of supporting the collective, and applied strategies to overcome the barriers. This persistence became especially apparent when contributors updated their reviews - often indicated through "update" or "addition" or "update + revision date" at the end of the review - based on positive enforcement of their motivation, e.g. after an app update. 

When re-assessing the app, common positive perceptions addressed the improvements app updates added, while negative perceptions were expressed through the errors experienced during the use. When re-assessing their motivations after initial use, participants positively highlighted the contribution of the CDA in the fight against the pandemic, while other users expressed doubts about the effectiveness of the data donation in regards to the pandemic mitigation at the same time.

We observed a strong sense of persistence, which was particularly noticeable in the context of negative emotions when facing errors. In many cases, users did not remove the CDA from their smartphone but rather explained their particular errors and requested the RKI to fix them. Users also often waited for multiple updates and resumed donating their data.

Another common approach of dealing with errors was through peer support. Users discovered workarounds and shared their solutions in the reviews or informed their fellow data donors in case a particular update fixed an issue, taking the role of \textit{community representatives}.
To do so, some returned to their reviews and, e.g. reported that it took a few days for them until the day counter started to work. In other cases, users suggested to check the CDA website, posted quotes from it or shared replies from the app support.
\begin{quote} 
\emph{I think the purpose of the app is great and I trust the data privacy. A clear 'Go' from me. From the RKI website\textsuperscript{2}: "Currently, in the app the counter for days donated is not always up to date. Due to the high load of the servers, the data retrieval is delayed.". (App Store, version 1.0.2, 5 stars)}
\end{quote}

Workarounds were also reported by users that initially were not able to contribute due to an unsupported device. For example, Huawei smartwatches were not directly supported in the first versions of the app. However, users explained that when connecting their smartwatch from an unsupported manufacturer to the Google Fit App they were able to donate. Since the Google Fit App can solely run on a smartphone without a connected wearable, it also allowed users not owning a fitness tracker or smartwatch wearable to connect to the CDA and still donate their data.
Reviewers also recommended to always ensure a charged device since the app would draw significant battery. 
A strategy to overcome the lack of feedback given by the app was to check the day counter as an indicator to whether the donation worked:
\begin{quote} 
\emph{Especially in the beginning, I frequently checked the app to make sure, does it count the days, hoping that would mean the data for the day were transmitted (P6).}
\end{quote}

Users' behaviours also demonstrated a lot of patience, especially regarding errors with the day counter. Although users were uncertain if their data was even donated, they continued using the CDA, hoping the issue would be resolved in future updates. Some users reported that the set up phase and registration took quite some time and even multiple restarts did not lead to a drop-out.
Furthermore, issues with the app were also seen in the context of the COVID-19 pandemic. The exceptional circumstances were considered and an understanding was expressed for issues during the early phase after publication. Fellow data donors were even asked to show more compassion due to the unprecedented times. This was expressed through statements like "Take it easy and drink some tea".
Nonetheless, some users threatened to uninstall the CDA and stop sharing their data if their reported errors were not resolved in the near future. This was particularly often expressed when users were uncertain about their donated data being received. 
However, a "second chance" was often granted in combination with a set countdown of a couple of days:
\begin{quote} 
\emph{Unfortunately, it turns out only at the very end [of the registration process] that it can not be connected with Samsung Health. But maybe this changes in the next few days. So, it will not be uninstalled yet.  (Play store, version 1.0.1, 2 stars)}
\end{quote}


The strong commitment to the project was also illustrated by users' patience in regards to the incompatibility and anticipation for their device to be supported soon. On that matter, users wished for a larger range of supported devices, since they hoped for a considerable increase in the number of data donations and, consequently, better chances for the project's success. 

The persistence of CDA users, observed particularly in these updated reviews and the related investment of time to get the app running properly in order to donate data, is particularly remarkable given the expectation of immediate availability in the context of digital technology, where long loading times of websites can already lead to dropouts. 

\section{Discussion}
In this research, we explored the lived experience of the CDA, a virtual citizen science projects to mitigate the negative effects of a global pandemic. Our results contribute insights about motivational processes of citizen scientists in times of crises. We observed that users of the CDA showed collective motivation to donate their data to support the community. The crisis experience led to participants staying persistent using the CDA, even so they were dealing with technical problems and waiting for updates. While we observed frustration of the citizen scientist when faced with problems, common conceptions were positive sentiments towards the idea of data donation.

\subsection{Motivations in Times of Crises}
In relation to our RQ1, which refers to the motivations for participation in citizen science in times of crises, we found that collective motives were the dominant factor to contribute. This is consistent with Raddick et al. and Curtis~\cite{raddick2013galaxy, curtis2015online}. Contrary to Rotman et al. ~\cite{rotman2012dynamic}, who propose that motivations are initially egoistic and change at pivotal points in the interaction with VCS, we did not observe that users’ main motives were egoistic throughout the contribution cycle. However, we hypothesise that this can partly be explained by the limited interaction possibilities provided by the CDA. In other words, due to limited interactions with the VCS app and other citizen scientists, it was less likely for the motivations to change during the cycle of engagement with the project. Our results extend the findings from Batson et al \cite{batson2002four} and Rotman et al. \cite{rotman2012dynamic}. Based on the work from Batson et al. \cite{batson2002four} there is usually a mix of motives present when citizens engage in communities and these motives, according to \cite{rotman2012dynamic}, change during the participation. 

However, our results showcased that this does not necessarily seem to be the case for VCS projects in times of crises. We hypothesise that the initial motivation of citizen scientist in such extraordinary circumstances might be strong enough to remain stable even if confronted with negative experiences such as a lack of interaction with the scientists and frustrating issues with the VCS app.


In regards to RQ2, the socio-psychological processes in VCS projects in times of crises, we found that peer support was common when dealing with problems of the app. Through providing workarounds and relaying information from the CDA website, some users indicated their commitment to the project. Besides the communication among volunteers, in contrast to other citizen science projects, there was very little to no communication between scientists and users. This is critical for VCS \cite{Reeves2019} and for the field of crisis informatics, for which our model of citizens’ motivations towards such apps has important implications. In line with Tan et al.~\cite{tan2017mobile}, who postulate a more citizen-centric communication in crises, we argue that serving the collective motives of participants will increase the understanding of the purpose of the project and the nature of their contributions. Some of the problems and questions raised in the reviews were answered by other users or covered in the FAQs on the CDA website. We assume that more effective communication could have reduced some of the frustration expressed in the reviews. Our results highlight the potential of supporting motivated participants to support their peers in using technologies such as the CDA. A practical implication could be to empower CDA users to support each other, which is motivated by our observation that peers adapted communication channels to communicate with each other that were not actually designed for such interactions. More precisely, citizens communicated about the intricacies of dealing with the CDA via online reviews in the Apple App Store and Google Play Store.

Factors deemed important for engagement in VCS, such as incorporating different motivational needs~\cite{Eveleigh2014Dabblers}, visualisations~\cite{Snyder2017}, gamification~\cite{Eveleigh2013}, and  responding to discussion messages~\cite{Reeves2019} seemed to be secondary for engagement in the CDA. Still, more than 500,000 people downloaded the app and connected their wearable device to donate data. One explanation for the high participation could be that the CDA app was promoted by the German government through public broadcasting television and various press reports. 

\subsection{Design Implications}
Based on our findings, we highlight possible ways forward for designing  VCS projects in crises contexts, as well as starting points that can help guiding the design of future systems in the area of crisis informatics.  
 
\subsubsection{Communicating Societal Relevance}
First, we address the importance of communicating societal relevance. The pandemic has already had drastic consequences on many societies worldwide \cite{johns2020covid}. Official authorities, such as the RKI, the German government and news outlets highlighted the role of the app in fighting the pandemic. We hypothesise that communicating the societal relevance of the CDA had an important effect on the rapid adoption rate of the app. From these observations we suggest that VCS projects in crises situations should be endorsed by official authorities and highlight their societal impact and communicate it clearly to the public. This resonates with findings from previous work. For instance, Gui et al~\cite{Gui2018} showed that people have a tendency of thinking beyond their individual needs in times of crisis. This effect, combined with the efforts of the authorities to communicate the societal relevance of the CDA may have supported the persistence of users.

\subsubsection{Cultivating Trust and Transparency}
Besides technical difficulties, concerns about data transparency were the main reason for users to stop using the CDA and contributing data. The high-level goal of the CDA, to reduce the spread of COVID-19 with health data from wearable devices, was clear to most users. However, the specifics of how their health data helped the RKI, e.g. to detect local fever outbreaks, remained unknown for many. In line with Huang et al.~\cite{Huang2015} we found that citizens were morel likely to share data the closer they were to a crisis physically and emotionally. However, the lack of understanding how their data was useful raised concerns about the value of their participation. Consequently, we argue that in VCS projects that involve passive sharing of sensitive data, \textit{users need fine-grained information about the value of their contribution}. Thus, it is not enough to provide information on a high-level but on an individual level instead, e.g. through daily status reports. 

The authority of the RKI as a federal German agency and the support from official authorities such as the German government led to a strong identification of the users with the goals of the CDA. Our results showed that ensuring data transparency and partnering with trusted institutions is essential to foster trust and long-term engagement of citizen scientists. Furthermore, contributors shared sensitive data with the RKI which they would not share with other platforms. While increased participation is essential for the success of VCS, developers acting on behalf of trusted institutions should be aware of their responsibility for the users who trust the institutions represented by the developers. Bowser et al.~\cite{bowser2017accounting} refer to this problematic priming on openness in the context of citizen science with suggesting that volunteers may not raise privacy concerns on their own.

\subsubsection{Fostering Community Support}
As people are often overwhelmed by the range of information available, especially during crises~\cite{Chauhan2017}, it is not enough to simply provide information. However, often, in crises, time and human resources to distribute important information are scarce. We found that many users were unaware of information that was available in the FAQ on the CDA website, for example that a non-functional day counter did not mean they were not donating data. Other topics frequently addressed in the reviews, e.g. reasons for excluding devices or data privacy policies, led to concerns about the integrity of the project. Identifying and clarifying such misconceptions could have prevented drop-outs. One way forward could be the launch of a designated platform to foster peer support such as a CDA forum directly integrated in the app. Another option could be the presentation of Twitter hashtags together with a VCS project to support community-based communication via social media~\cite{starbird2011voluntweeters, Vieweg2010}. Previous work has shown that digital volunteers in crisis situations use Twitter to gather and disperse information and that such information is relevant, accurate and useful \cite{Vieweg2010}. Online platforms such as Twitter and other internet forums support disaster-related citizen participation, enabling individual capacities and collective action \cite{starbird2011voluntweeters}. Consistent with these observations, we propose that such platforms could engage \textit{volunteers who act as community representatives} for the scientists to mitigate some of the communication issues we observed in the current project, for example. As a positive side effect, these volunteers take ownership of the project, possibly synthesising user feedback for the developers.

\subsubsection{Personal Crisis Informatics}
Previous studies in crisis informatics primarily focused on analysing how different social media platforms, e.g. Reddit or Twitter, were used in times of crises, e.g.~\cite{Gui2018}. However, there is unexplored potential at the intersection of VCS and crisis informatics. For instance, the design and analysis of the CDA can inspire novel crisis informatics systems beyond social media. In other words, systems similar to the CDA could be an interesting extension of previous crisis informatics approaches.

The CDA is unlike more traditional VCS projects. It collects crisis-relevant, sensitive, personal data using personal informatics technologies which are normally used for private purposes, e.g. fitness trackers. Interestingly, we observed that citizens were willing to share not only their sensitive data (e.g. resting heart rate) requested by the CDA, but also broader metrics not requested by the system, e.g. body temperature.

Our analysis showed that the participants exhibited a high level of willingness to share sensitive data for a social purpose. Fostering such willingness appears to be a key design goal for technologies which use data at the time of crisis. Citizens communicated that they would have liked to actively log additional data such as their daily temperature to support the RKI and they did not understand why it was not possible to do so. Furthermore, citizens voiced their frustration with the CDA supporting only specific wearables and, which rendered them unable to donate their data. Future systems in the area of crisis informatics could take advantage of this readiness to help and provide a broader set of systems and platforms for people with different levels of openness towards technology and technology self-efficacy. This suggests that, in future public health crises, using existing health apps for data donations together with apps designed specifically for the crisis (similar to the CDA) can render the best results. In addition, a simple website could be designed on which people who do not own a smartphone or wearable could donate their data through manual input.

\subsection{Generalisability}
Users considered the CDA as an opportunity to serve for the greater good and this perception was further reinforced by media coverage and the RKI's role as a government organisation. We argue that our findings can be generalised to projects that are deployed within comparable circumstances, e.g. official authorities fighting large scale public health crises. For example, authorities could launch similar services to monitor outbreaks of dengue fever, measles or influenza. For the control of dengue fever, it is essential to reduce mosquito populations, as there are no effective vaccines \cite{NatureDengue}. The recommended management strategy is to eliminate unnecessary container habitats that collect water. Another public health hazard are measles, a highly contagious, serious disease caused by a virus \cite{WHOmeasles}. Even though there is an effective vaccine, in 2018, there were more than 140 000 measles deaths globally, especially in developing countries. For both diseases, surveillance of fever curves in conjunction with manual reporting of infections might contribute to the localisation of outbreaks and help authorities in narrowing down their efforts to regions with high prevalence. Given the technical difficulties that accompanied the launch of the CDA, one approach for health authorities in future epidemics could be to officially endorse existing providers, such as the Fitbit platform. Users could opt-in to make their data available for the purpose of detecting symptoms based on their fitness tracker data, which has been shown to work to predict influenza-like illnesses \cite{radin2020harnessing}. With this method, an established technical infrastructure could be adopted by a trusted organisation to ensure reliable measurements in large populations.

In the context of a pandemic, we assume that that the high perceived relevance of the projects addresses users' collective motivations and makes them accept a range of adversities, which we will outline in the following. First, in contrast to many crisis informatics projects, where participants share observations, relay information or map hazards, the CDA requires users to contribute sensitive personal data, such as heart rate and activity data. Second, not only were there no direct rewards or a gratification system in place as a compensation for using the CDA, nor did the CDA provide clear information on how exactly the data were useful in mitigating the effects of the pandemic. Lastly, participants shared data continuously with no defined end date, which is different to episodic crises, where citizen scientists often provide only a few data points, e.g. extent of damage at a certain location over a limited amount of time.

Therefore, the CDA shows that the societal relevance of the crisis might overrule common inhibitors for participation in VCS and crisis informatics projects. Our findings suggest that users in different types of crises with lower perceived relevance might have stopped engaging with the project at earlier stages already. Users potentially accept adversities should they have the feeling they can contribute to the greater good. In this case, developers could address users' collective motivations and express their appreciation for serving the community. For example, the app could acknowledge the donation and send participants thank you' notifications or let them share a message about their contributions on social media. However, we want to highlight that more research is needed to determine what drives citizen scientists' participation during different public health crises.

\subsection{Limitations}
While we aimed to make a contribution towards understanding the lived experience of a virtual citizen science project during the COVID-19 pandemic, we can identify some limitations in our work. To this day, there is, to the best of our knowledge, no evidence that the approach of using physiological data from fitness trackers to predict COVID-19 hot-spots is effective. Moreover, most efforts to utilise mobile devices for the mitigation of the pandemic focus on contact tracing apps. These apps do not require to share personal data with the providers. Therefore, the relevance of our work for the current pandemic might be limited and is more significant for future VCS projects in crisis contexts that rely on personal data. 

Our results are largely based on the analysis of reviews from the Apple App Store and Google Play Store. We are aware that most app store reviews are rather extreme \cite{hu2006can}, being very positive or very negative. Additionally, future work needs to determine if our interview sample reflects the population of CDA users in terms of its demographics. However, our case-study presents an in-depth account of attitudes and behaviours in a nuanced manner, as it distinguishes between positive and negative emotional responses to the CDA, e.g. drop-outs or continued contribution. For a more generaliseable understanding of the identified motivations, for instance in similar public health crises, future work could juxtapose the results of a more large-scale study with our findings.

The CDA was developed in a short time frame due to the dynamic of the pandemic and the urgent need for measures to detect infections and isolate potential carriers of the virus as fast as possible. Due to the time constraints, the app was released with several severe limitations, mainly the non-functional day counter, a limited range of supported devices and a lack of information on how the donated data was used by the RKI. These shortcomings could potentially have been prevented through a user-centred design process and systematic user testing. For future crisis VCS or crisis informatics projects that face the same constraints, we suggest that even when there is not enough time for user involvement in the initial releases, developers provide and maintain social media channels to synthesise feedback for future iterations and give users an opportunity to voice their opinion. This is consistent with the recommendation to utilise social media for communication among participants and between participants and developers. The negative reviews that were associated with these shortcomings, contributed to our observation of persistence, i.e. the continued participation even after experiencing problems. Had the CDA been released without these problems, persistence may not have been among our key findings. However, some of our other findings, such as the willingness to serve for the greater good, similarly support the prevalence of collective motives.


\section{Conclusion}
In this paper we studied the motivations of citizen scientists to contribute to a VCS project in a time of crises (COVID-19 pandemic). For doing so, we collected and analysed reviews of the Corona Data-Donation App, a mobile application released by the RKI in Germany. This application allows users to donate their physiological data using wearables, in order to mitigate the effects of the pandemic by possibly detecting local fever outbreaks that could be associated with COVID-19. 
We then conducted additional in-depth interviews with individuals that had experience with using this application, and analysed both the reviews and interviews. We thus could draw a picture of the users' lived experience of participating in this VCS project, illustrating the cycle of engagement of the citizen scientists and people's motivations to contribute such personal data under these circumstances.
Our contribution in this paper is fourfold: First, we reveal that collective motives prevailed over egoistic motives driving the citizens to contribute to the CDA. Second, the emotional response of the users included both positive and negative emotions, and we demonstrated which aspects of using the CDA resulted in particular emotions. We also show the repercussions this had on the overall use of the app and continuation of contributing to the VCS project. Third, we uncovered that citizen scientists of this particular project demonstrated a very strong sense of persistence, i.e. a large proportion continued to use the app in spite of various issues "for the greater good". 
Finally, we present implications for the design and presentation of future VCS projects, highlighting the importance of communication, transparency and responsibility. We thus drew insights for what motivates people to participate in VCS projects in times of crises and how designers of such systems can hopefully benefit from our findings in the future.










\begin{acks}
We thank the reviewers of this paper for their time and the valuable feedback they provided on previous versions of this manuscript. 
This research was supported in parts by a Lichtenbergprofessorship of the Volkswagenfoundation, the BMWi funded network KI-SIGS (grant 01MK20012), the BMBF project InviDas (grant 16SV8539) and the Leibniz ScienceCampus Bremen Digital Public Health, which is jointly funded by the Leibniz Association (W4/2018), the Federal State of Bremen and the Leibniz Institute for Prevention Research and Epidemiology – BIPS.
\end{acks}

\bibliographystyle{ACM-Reference-Format}
\bibliography{BibSharingHearts}


\begin{thebibliography}{98}


\ifx \showCODEN    \undefined \def \showCODEN     #1{\unskip}     \fi
\ifx \showDOI      \undefined \def \showDOI       #1{#1}\fi
\ifx \showISBNx    \undefined \def \showISBNx     #1{\unskip}     \fi
\ifx \showISBNxiii \undefined \def \showISBNxiii  #1{\unskip}     \fi
\ifx \showISSN     \undefined \def \showISSN      #1{\unskip}     \fi
\ifx \showLCCN     \undefined \def \showLCCN      #1{\unskip}     \fi
\ifx \shownote     \undefined \def \shownote      #1{#1}          \fi
\ifx \showarticletitle \undefined \def \showarticletitle #1{#1}   \fi
\ifx \showURL      \undefined \def \showURL       {\relax}        \fi
\providecommand\bibfield[2]{#2}
\providecommand\bibinfo[2]{#2}
\providecommand\natexlab[1]{#1}
\providecommand\showeprint[2][]{arXiv:#2}

\bibitem[\protect\citeauthoryear{Batson, Ahmad, and Tsang}{Batson
  et~al\mbox{.}}{2002}]%
        {batson2002four}
\bibfield{author}{\bibinfo{person}{Daniel Batson}, \bibinfo{person}{Nadia
  Ahmad}, {and} \bibinfo{person}{Jo-Ann Tsang}.}
  \bibinfo{year}{2002}\natexlab{}.
\newblock \showarticletitle{Four motives for community involvement}.
\newblock \bibinfo{journal}{\emph{Journal of social issues}}
  \bibinfo{volume}{58}, \bibinfo{number}{3} (\bibinfo{year}{2002}),
  \bibinfo{pages}{429--445}.
\newblock


\bibitem[\protect\citeauthoryear{Blandford, Furniss, and Makri}{Blandford
  et~al\mbox{.}}{2016}]%
        {blandford_qualitative_2016}
\bibfield{author}{\bibinfo{person}{Ann Blandford}, \bibinfo{person}{Dominic
  Furniss}, {and} \bibinfo{person}{Stephann Makri}.}
  \bibinfo{year}{2016}\natexlab{}.
\newblock \showarticletitle{Qualitative {HCI} {Research}: {Going} {Behind} the
  {Scenes}}.
\newblock \bibinfo{journal}{\emph{Synthesis Lectures on Human-Centered
  Informatics}} \bibinfo{volume}{9}, \bibinfo{number}{1} (\bibinfo{date}{April}
  \bibinfo{year}{2016}), \bibinfo{pages}{1--115}.
\newblock
\showISSN{1946-7680}
\urldef\tempurl%
\url{https://doi.org/10.2200/S00706ED1V01Y201602HCI034}
\showDOI{\tempurl}
\newblock
\shownote{Publisher: Morgan \& Claypool Publishers.}


\bibitem[\protect\citeauthoryear{Bowser, Shilton, Preece, and Warrick}{Bowser
  et~al\mbox{.}}{2017}]%
        {bowser2017accounting}
\bibfield{author}{\bibinfo{person}{Anne Bowser}, \bibinfo{person}{Katie
  Shilton}, \bibinfo{person}{Jenny Preece}, {and} \bibinfo{person}{Elizabeth
  Warrick}.} \bibinfo{year}{2017}\natexlab{}.
\newblock \showarticletitle{Accounting for privacy in citizen science: Ethical
  research in a context of openness}. In \bibinfo{booktitle}{\emph{Proceedings
  of the 2017 ACM Conference on Computer Supported Cooperative Work and Social
  Computing}}. \bibinfo{pages}{2124--2136}.
\newblock


\bibitem[\protect\citeauthoryear{Boyd, Golder, and Lotan}{Boyd
  et~al\mbox{.}}{2010}]%
        {boyd2010tweet}
\bibfield{author}{\bibinfo{person}{Danah Boyd}, \bibinfo{person}{Scott Golder},
  {and} \bibinfo{person}{Gilad Lotan}.} \bibinfo{year}{2010}\natexlab{}.
\newblock \showarticletitle{Tweet, tweet, retweet: Conversational aspects of
  retweeting on twitter}. In \bibinfo{booktitle}{\emph{2010 43rd Hawaii
  international conference on system sciences}}. IEEE, \bibinfo{pages}{1--10}.
\newblock


\bibitem[\protect\citeauthoryear{Brown and Juhlin}{Brown and Juhlin}{2015}]%
        {brown2015enjoying}
\bibfield{author}{\bibinfo{person}{Barry Brown} {and} \bibinfo{person}{Oskar
  Juhlin}.} \bibinfo{year}{2015}\natexlab{}.
\newblock \bibinfo{booktitle}{\emph{Enjoying machines}}.
\newblock \bibinfo{publisher}{Mit Press}.
\newblock


\bibitem[\protect\citeauthoryear{Busvine}{Busvine}{2020}]%
        {busvine_2020}
\bibfield{author}{\bibinfo{person}{Douglas Busvine}.}
  \bibinfo{year}{2020}\natexlab{}.
\newblock \bibinfo{booktitle}{\emph{Germany launches smartwatch app to monitor
  coronavirus spread}}.
\newblock
\urldef\tempurl%
\url{https://www.reuters.com/article/us-health-coronavirus-germany-tech/germany-launches-smartwatch-app-to-monitor-coronavirus-spread-idUSKBN21P1SS}
\showURL{%
Retrieved September 14, 2020 from \tempurl}


\bibitem[\protect\citeauthoryear{Butler, Sproull, Kiesler, and Kraut}{Butler
  et~al\mbox{.}}{2002}]%
        {butler2002community}
\bibfield{author}{\bibinfo{person}{Brian Butler}, \bibinfo{person}{Lee
  Sproull}, \bibinfo{person}{Sara Kiesler}, {and} \bibinfo{person}{Robert
  Kraut}.} \bibinfo{year}{2002}\natexlab{}.
\newblock \showarticletitle{Community effort in online groups: Who does the
  work and why}.
\newblock \bibinfo{journal}{\emph{Leadership at a distance: Research in
  technologically supported work}}  \bibinfo{volume}{1} (\bibinfo{year}{2002}),
  \bibinfo{pages}{171--194}.
\newblock


\bibitem[\protect\citeauthoryear{{ccc.de}}{{ccc.de}}{2020}]%
        {CCCdatenspende}
\bibfield{author}{\bibinfo{person}{{ccc.de}}.} \bibinfo{year}{2020}\natexlab{}.
\newblock \bibinfo{booktitle}{\emph{CCC analysiert Corona-Datenspende des
  RKI}}.
\newblock
\urldef\tempurl%
\url{https://www.ccc.de/de/updates/2020/abofalle-datenspende}
\showURL{%
Retrieved December 17, 2020 from \tempurl}


\bibitem[\protect\citeauthoryear{CEBM}{CEBM}{2020}]%
        {cebm_2020}
\bibfield{author}{\bibinfo{person}{CEBM}.} \bibinfo{year}{2020}\natexlab{}.
\newblock \bibinfo{booktitle}{\emph{COVID-19: What proportion are
  asymptomatic?}}
\newblock
\urldef\tempurl%
\url{https://www.cebm.net/covid-19/covid-19-what-proportion-are-asymptomatic/}
\showURL{%
Retrieved September 14, 2020 from \tempurl}


\bibitem[\protect\citeauthoryear{{Centers for Disease Control and Prevention,
  USA}}{{Centers for Disease Control and Prevention, USA}}{2020}]%
        {cdc_coronavirus_2020}
\bibfield{author}{\bibinfo{person}{{Centers for Disease Control and Prevention,
  USA}}.} \bibinfo{year}{2020}\natexlab{}.
\newblock \bibinfo{booktitle}{\emph{Coronavirus {Disease} 2019 ({COVID}-19) -
  {Transmission}}}.
\newblock
\urldef\tempurl%
\url{https://www.cdc.gov/coronavirus/2019-ncov/prevent-getting-sick/how-covid-spreads.html}
\showURL{%
Retrieved September 14, 2020 from \tempurl}


\bibitem[\protect\citeauthoryear{{Charite, Germany}}{{Charite,
  Germany}}{2020}]%
        {covapp2020}
\bibfield{author}{\bibinfo{person}{{Charite, Germany}}.}
  \bibinfo{year}{2020}\natexlab{}.
\newblock \bibinfo{booktitle}{\emph{CovApp: Recommendations and information
  regarding coronavirus}}.
\newblock
\urldef\tempurl%
\url{https://covapp.charite.de/}
\showURL{%
Retrieved September 14, 2020 from \tempurl}


\bibitem[\protect\citeauthoryear{Chauhan and Hughes}{Chauhan and
  Hughes}{2017}]%
        {Chauhan2017}
\bibfield{author}{\bibinfo{person}{Apoorva Chauhan} {and}
  \bibinfo{person}{Amanda~L. Hughes}.} \bibinfo{year}{2017}\natexlab{}.
\newblock \showarticletitle{Providing Online Crisis Information: An Analysis of
  Official Sources during the 2014 Carlton Complex Wildfire}. In
  \bibinfo{booktitle}{\emph{Proceedings of the 2017 CHI Conference on Human
  Factors in Computing Systems}} (Denver, Colorado, USA)
  \emph{(\bibinfo{series}{CHI '17})}. \bibinfo{publisher}{Association for
  Computing Machinery}, \bibinfo{address}{New York, NY, USA},
  \bibinfo{pages}{3151–3162}.
\newblock
\showISBNx{9781450346559}
\urldef\tempurl%
\url{https://doi.org/10.1145/3025453.3025627}
\showDOI{\tempurl}


\bibitem[\protect\citeauthoryear{Cranshaw and Kittur}{Cranshaw and
  Kittur}{2011}]%
        {cranshaw2011polymath}
\bibfield{author}{\bibinfo{person}{Justin Cranshaw} {and}
  \bibinfo{person}{Aniket Kittur}.} \bibinfo{year}{2011}\natexlab{}.
\newblock \showarticletitle{The polymath project: lessons from a successful
  online collaboration in mathematics}. In
  \bibinfo{booktitle}{\emph{Proceedings of the SIGCHI conference on human
  factors in computing systems}}. \bibinfo{pages}{1865--1874}.
\newblock


\bibitem[\protect\citeauthoryear{Curtis}{Curtis}{2015}]%
        {curtis2015online}
\bibfield{author}{\bibinfo{person}{Vickie Curtis}.}
  \bibinfo{year}{2015}\natexlab{}.
\newblock \emph{\bibinfo{title}{Online citizen science projects: an exploration
  of motivation, contribution and participation}}.
\newblock \bibinfo{thesistype}{Ph.D. Dissertation}. \bibinfo{school}{The Open
  University}.
\newblock


\bibitem[\protect\citeauthoryear{Davalbhakta, Advani, Kumar, Agarwal, Bhoyar,
  Fedirko, Misra, Goel, Gupta, and Agarwal}{Davalbhakta et~al\mbox{.}}{2020}]%
        {davalbhakta2020systematic}
\bibfield{author}{\bibinfo{person}{Samira Davalbhakta},
  \bibinfo{person}{Shailesh Advani}, \bibinfo{person}{Shobhit Kumar},
  \bibinfo{person}{Vishwesh Agarwal}, \bibinfo{person}{Samruddhi Bhoyar},
  \bibinfo{person}{Elizabeth Fedirko}, \bibinfo{person}{Durga Misra},
  \bibinfo{person}{Ashish Goel}, \bibinfo{person}{Latika Gupta}, {and}
  \bibinfo{person}{Vikas Agarwal}.} \bibinfo{year}{2020}\natexlab{}.
\newblock \showarticletitle{A systematic review of the smartphone applications
  available for coronavirus disease 2019 (COVID19) and their assessment using
  the mobile app rating scale (MARS)}.
\newblock \bibinfo{journal}{\emph{medRxiv}} (\bibinfo{year}{2020}).
\newblock


\bibitem[\protect\citeauthoryear{Dong, Du, and Gardner}{Dong
  et~al\mbox{.}}{2020}]%
        {dong2020interactive}
\bibfield{author}{\bibinfo{person}{Ensheng Dong}, \bibinfo{person}{Hongru Du},
  {and} \bibinfo{person}{Lauren Gardner}.} \bibinfo{year}{2020}\natexlab{}.
\newblock \showarticletitle{An interactive web-based dashboard to track
  COVID-19 in real time}.
\newblock \bibinfo{journal}{\emph{The Lancet infectious diseases}}
  \bibinfo{volume}{20}, \bibinfo{number}{5} (\bibinfo{year}{2020}),
  \bibinfo{pages}{533--534}.
\newblock


\bibitem[\protect\citeauthoryear{Doyle, David, Li, Luczak-Roesch, Anderson, and
  Pierson}{Doyle et~al\mbox{.}}{2019}]%
        {doyle_using_2019}
\bibfield{author}{\bibinfo{person}{Cathal Doyle}, \bibinfo{person}{Rodreck
  David}, \bibinfo{person}{Yevgeniya Li}, \bibinfo{person}{Markus
  Luczak-Roesch}, \bibinfo{person}{Dayle Anderson}, {and}
  \bibinfo{person}{Cameron~M. Pierson}.} \bibinfo{year}{2019}\natexlab{}.
\newblock \showarticletitle{Using the web for science in the classroom:
  {Online} citizen science participation in teaching and learning}. In
  \bibinfo{booktitle}{\emph{Proceedings of the 10th {ACM} {Conference} on {Web}
  {Science}}}. \bibinfo{pages}{71--80}.
\newblock


\bibitem[\protect\citeauthoryear{Education}{Education}{2014}]%
        {NatureDengue}
\bibfield{author}{\bibinfo{person}{Nature Education}.}
  \bibinfo{year}{2014}\natexlab{}.
\newblock \bibinfo{booktitle}{\emph{Controlling Dengue Outbreaks}}.
\newblock
\urldef\tempurl%
\url{https://www.nature.com/scitable/topicpage/controlling-dengue-outbreaks-22403714/}
\showURL{%
Retrieved December 28, 2020 from \tempurl}


\bibitem[\protect\citeauthoryear{Eitzel, Cappadonna, Santos-Lang, Duerr,
  Virapongse, West, Kyba, Bowser, Cooper, Sforzi, et~al\mbox{.}}{Eitzel
  et~al\mbox{.}}{2017}]%
        {eitzel2017citizen}
\bibfield{author}{\bibinfo{person}{Melissa~V Eitzel},
  \bibinfo{person}{Jessica~L Cappadonna}, \bibinfo{person}{Chris Santos-Lang},
  \bibinfo{person}{Ruth~Ellen Duerr}, \bibinfo{person}{Arika Virapongse},
  \bibinfo{person}{Sarah~Elizabeth West}, \bibinfo{person}{Christopher Kyba},
  \bibinfo{person}{Anne Bowser}, \bibinfo{person}{Caren~Beth Cooper},
  \bibinfo{person}{Andrea Sforzi}, {et~al\mbox{.}}}
  \bibinfo{year}{2017}\natexlab{}.
\newblock \showarticletitle{Citizen science terminology matters: exploring key
  terms}.
\newblock \bibinfo{journal}{\emph{Citizen science: Theory and practice}}
  \bibinfo{volume}{2}, \bibinfo{number}{1} (\bibinfo{year}{2017}).
\newblock


\bibitem[\protect\citeauthoryear{Epstein, Caraway, Johnston, Ping, Fogarty, and
  Munson}{Epstein et~al\mbox{.}}{2016}]%
        {Epstein2016}
\bibfield{author}{\bibinfo{person}{Daniel~A. Epstein}, \bibinfo{person}{Monica
  Caraway}, \bibinfo{person}{Chuck Johnston}, \bibinfo{person}{An Ping},
  \bibinfo{person}{James Fogarty}, {and} \bibinfo{person}{Sean~A. Munson}.}
  \bibinfo{year}{2016}\natexlab{}.
\newblock \showarticletitle{Beyond Abandonment to Next Steps: Understanding and
  Designing for Life after Personal Informatics Tool Use}. In
  \bibinfo{booktitle}{\emph{Proceedings of the 2016 CHI Conference on Human
  Factors in Computing Systems}} (San Jose, California, USA)
  \emph{(\bibinfo{series}{CHI '16})}. \bibinfo{publisher}{Association for
  Computing Machinery}, \bibinfo{address}{New York, NY, USA},
  \bibinfo{pages}{1109–1113}.
\newblock
\showISBNx{9781450333627}
\urldef\tempurl%
\url{https://doi.org/10.1145/2858036.2858045}
\showDOI{\tempurl}


\bibitem[\protect\citeauthoryear{Epstein, Ping, Fogarty, and Munson}{Epstein
  et~al\mbox{.}}{2015}]%
        {Epstein2015}
\bibfield{author}{\bibinfo{person}{Daniel~A. Epstein}, \bibinfo{person}{An
  Ping}, \bibinfo{person}{James Fogarty}, {and} \bibinfo{person}{Sean~A.
  Munson}.} \bibinfo{year}{2015}\natexlab{}.
\newblock \showarticletitle{A Lived Informatics Model of Personal Informatics}.
  In \bibinfo{booktitle}{\emph{Proceedings of the 2015 ACM International Joint
  Conference on Pervasive and Ubiquitous Computing}} (Osaka, Japan)
  \emph{(\bibinfo{series}{UbiComp '15})}. \bibinfo{publisher}{Association for
  Computing Machinery}, \bibinfo{address}{New York, NY, USA},
  \bibinfo{pages}{731–742}.
\newblock
\showISBNx{9781450335744}
\urldef\tempurl%
\url{https://doi.org/10.1145/2750858.2804250}
\showDOI{\tempurl}


\bibitem[\protect\citeauthoryear{{European Cluster Collaboration
  Platform}}{{European Cluster Collaboration Platform}}{2020}]%
        {ProteGOSafe}
\bibfield{author}{\bibinfo{person}{{European Cluster Collaboration Platform}}.}
  \bibinfo{year}{2020}\natexlab{}.
\newblock \bibinfo{booktitle}{\emph{ProteGO Safe application}}.
\newblock
\urldef\tempurl%
\url{https://www.clustercollaboration.eu/forum/best-practices-sharing/protego-safe-application}
\showURL{%
Retrieved September 14, 2020 from \tempurl}


\bibitem[\protect\citeauthoryear{Eveleigh, Jennett, Blandford, Brohan, and
  Cox}{Eveleigh et~al\mbox{.}}{2014}]%
        {Eveleigh2014Dabblers}
\bibfield{author}{\bibinfo{person}{Alexandra Eveleigh},
  \bibinfo{person}{Charlene Jennett}, \bibinfo{person}{Ann Blandford},
  \bibinfo{person}{Philip Brohan}, {and} \bibinfo{person}{Anna~L. Cox}.}
  \bibinfo{year}{2014}\natexlab{}.
\newblock \showarticletitle{Designing for Dabblers and Deterring Drop-Outs in
  Citizen Science}. In \bibinfo{booktitle}{\emph{Proceedings of the SIGCHI
  Conference on Human Factors in Computing Systems}} (Toronto, Ontario, Canada)
  \emph{(\bibinfo{series}{CHI '14})}. \bibinfo{publisher}{Association for
  Computing Machinery}, \bibinfo{address}{New York, NY, USA},
  \bibinfo{pages}{2985–2994}.
\newblock
\showISBNx{9781450324731}
\urldef\tempurl%
\url{https://doi.org/10.1145/2556288.2557262}
\showDOI{\tempurl}


\bibitem[\protect\citeauthoryear{Eveleigh, Jennett, Lynn, and Cox}{Eveleigh
  et~al\mbox{.}}{2013}]%
        {Eveleigh2013}
\bibfield{author}{\bibinfo{person}{Alexandra Eveleigh},
  \bibinfo{person}{Charlene Jennett}, \bibinfo{person}{Stuart Lynn}, {and}
  \bibinfo{person}{Anna~L. Cox}.} \bibinfo{year}{2013}\natexlab{}.
\newblock \showarticletitle{“I Want to Be a Captain! I Want to Be a
  Captain!”: Gamification in the Old Weather Citizen Science Project}. In
  \bibinfo{booktitle}{\emph{Proceedings of the First International Conference
  on Gameful Design, Research, and Applications}} (Toronto, Ontario, Canada)
  \emph{(\bibinfo{series}{Gamification '13})}. \bibinfo{publisher}{Association
  for Computing Machinery}, \bibinfo{address}{New York, NY, USA},
  \bibinfo{pages}{79–82}.
\newblock
\showISBNx{9781450328159}
\urldef\tempurl%
\url{https://doi.org/10.1145/2583008.2583019}
\showDOI{\tempurl}


\bibitem[\protect\citeauthoryear{{Federal Press Office Germany}}{{Federal Press
  Office Germany}}{2020}]%
        {bundesregierung2020datenspende}
\bibfield{author}{\bibinfo{person}{{Federal Press Office Germany}}.}
  \bibinfo{year}{2020}\natexlab{}.
\newblock \bibinfo{booktitle}{\emph{Datenspende gegen Corona}}.
\newblock
\urldef\tempurl%
\url{https://www.bundesregierung.de/breg-de/themen/coronavirus/datenspende-app-1739928}
\showURL{%
Retrieved September 14, 2020 from \tempurl}


\bibitem[\protect\citeauthoryear{Ghosh, Badillo-Urquiola, Guha, LaViola~Jr, and
  Wisniewski}{Ghosh et~al\mbox{.}}{2018}]%
        {ghosh2018parental}
\bibfield{author}{\bibinfo{person}{Arup~Kumar Ghosh}, \bibinfo{person}{Karla
  Badillo-Urquiola}, \bibinfo{person}{Shion Guha}, \bibinfo{person}{Joseph~J.
  LaViola~Jr}, {and} \bibinfo{person}{Pamela~J. Wisniewski}.}
  \bibinfo{year}{2018}\natexlab{}.
\newblock \showarticletitle{Safety vs. Surveillance: What Children Have to Say
  about Mobile Apps for Parental Control}. In
  \bibinfo{booktitle}{\emph{Proceedings of the 2018 CHI Conference on Human
  Factors in Computing Systems}} (Montreal QC, Canada)
  \emph{(\bibinfo{series}{CHI '18})}. \bibinfo{publisher}{Association for
  Computing Machinery}, \bibinfo{address}{New York, NY, USA},
  \bibinfo{pages}{1–14}.
\newblock
\showISBNx{9781450356206}
\urldef\tempurl%
\url{https://doi.org/10.1145/3173574.3173698}
\showDOI{\tempurl}


\bibitem[\protect\citeauthoryear{Gui, Kou, Pine, Ladaw, Kim, Suzuki-Gill, and
  Chen}{Gui et~al\mbox{.}}{2018}]%
        {Gui2018}
\bibfield{author}{\bibinfo{person}{Xinning Gui}, \bibinfo{person}{Yubo Kou},
  \bibinfo{person}{Kathleen Pine}, \bibinfo{person}{Elisa Ladaw},
  \bibinfo{person}{Harold Kim}, \bibinfo{person}{Eli Suzuki-Gill}, {and}
  \bibinfo{person}{Yunan Chen}.} \bibinfo{year}{2018}\natexlab{}.
\newblock \showarticletitle{Multidimensional Risk Communication: Public
  Discourse on Risks during an Emerging Epidemic}. In
  \bibinfo{booktitle}{\emph{Proceedings of the 2018 CHI Conference on Human
  Factors in Computing Systems}} (Montreal QC, Canada)
  \emph{(\bibinfo{series}{CHI '18})}. \bibinfo{publisher}{Association for
  Computing Machinery}, \bibinfo{address}{New York, NY, USA},
  \bibinfo{pages}{1–14}.
\newblock
\showISBNx{9781450356206}
\urldef\tempurl%
\url{https://doi.org/10.1145/3173574.3173788}
\showDOI{\tempurl}


\bibitem[\protect\citeauthoryear{Gura}{Gura}{2013}]%
        {gura_citizen_2013}
\bibfield{author}{\bibinfo{person}{Trisha Gura}.}
  \bibinfo{year}{2013}\natexlab{}.
\newblock \showarticletitle{Citizen science: amateur experts}.
\newblock \bibinfo{journal}{\emph{Nature}} \bibinfo{volume}{496},
  \bibinfo{number}{7444} (\bibinfo{year}{2013}), \bibinfo{pages}{259--261}.
\newblock
\newblock
\shownote{ISBN: 0028-0836 Publisher: Nature Research.}


\bibitem[\protect\citeauthoryear{Hagar}{Hagar}{2006}]%
        {hagar2006using}
\bibfield{author}{\bibinfo{person}{Chris Hagar}.}
  \bibinfo{year}{2006}\natexlab{}.
\newblock \showarticletitle{Using research to aid the design of a crisis
  information management course}. In \bibinfo{booktitle}{\emph{ALISE Annual
  Conference SIG Multicultural, Ethnic \& Humanistic Concerns (MEH).
  Information Seeking and Service Delivery for Communities in Disaster/Crisis,
  San Antonio}}.
\newblock


\bibitem[\protect\citeauthoryear{Hamilton}{Hamilton}{2020}]%
        {hamilton_2020}
\bibfield{author}{\bibinfo{person}{Isobel~Asher Hamilton}.}
  \bibinfo{year}{2020}\natexlab{}.
\newblock \bibinfo{booktitle}{\emph{Poland made an app that forces coronavirus
  patients to take regular selfies to prove they're indoors or face a police
  visit}}.
\newblock
\urldef\tempurl%
\url{https://www.businessinsider.com/poland-app-coronavirus-patients-mandaotory-selfie-2020-3?r=DE&IR=T}
\showURL{%
Retrieved September 14, 2020 from \tempurl}


\bibitem[\protect\citeauthoryear{Hand}{Hand}{2010}]%
        {hand_people_2010}
\bibfield{author}{\bibinfo{person}{Eric Hand}.}
  \bibinfo{year}{2010}\natexlab{}.
\newblock \showarticletitle{People power: networks of human minds are taking
  citizen science to a new level}.
\newblock \bibinfo{journal}{\emph{Nature}} \bibinfo{volume}{466},
  \bibinfo{number}{7307} (\bibinfo{year}{2010}), \bibinfo{pages}{685--688}.
\newblock
\newblock
\shownote{ISBN: 0028-0836 Publisher: Nature Publishing Group.}


\bibitem[\protect\citeauthoryear{Hedegaard and Simonsen}{Hedegaard and
  Simonsen}{2013}]%
        {hedegaard2013extracting}
\bibfield{author}{\bibinfo{person}{Steffen Hedegaard} {and}
  \bibinfo{person}{Jakob~Grue Simonsen}.} \bibinfo{year}{2013}\natexlab{}.
\newblock \showarticletitle{Extracting usability and user experience
  information from online user reviews}. In
  \bibinfo{booktitle}{\emph{Proceedings of the SIGCHI Conference on Human
  Factors in Computing Systems}}. \bibinfo{pages}{2089--2098}.
\newblock


\bibitem[\protect\citeauthoryear{Hertel, Niedner, and Herrmann}{Hertel
  et~al\mbox{.}}{2003}]%
        {hertel2003motivation}
\bibfield{author}{\bibinfo{person}{Guido Hertel}, \bibinfo{person}{Sven
  Niedner}, {and} \bibinfo{person}{Stefanie Herrmann}.}
  \bibinfo{year}{2003}\natexlab{}.
\newblock \showarticletitle{Motivation of software developers in Open Source
  projects: an Internet-based survey of contributors to the Linux kernel}.
\newblock \bibinfo{journal}{\emph{Research policy}} \bibinfo{volume}{32},
  \bibinfo{number}{7} (\bibinfo{year}{2003}), \bibinfo{pages}{1159--1177}.
\newblock


\bibitem[\protect\citeauthoryear{Hu, Pavlou, and Zhang}{Hu
  et~al\mbox{.}}{2006}]%
        {hu2006can}
\bibfield{author}{\bibinfo{person}{Nan Hu}, \bibinfo{person}{Paul~A Pavlou},
  {and} \bibinfo{person}{Jennifer Zhang}.} \bibinfo{year}{2006}\natexlab{}.
\newblock \showarticletitle{Can online reviews reveal a product's true quality?
  Empirical findings and analytical modeling of online word-of-mouth
  communication}. In \bibinfo{booktitle}{\emph{Proceedings of the 7th ACM
  conference on Electronic commerce}}. \bibinfo{pages}{324--330}.
\newblock


\bibitem[\protect\citeauthoryear{Huang, Wang, Li, Ren, Zhao, Hu, Zhang, Fan,
  Xu, and Gu}{Huang et~al\mbox{.}}{2020}]%
        {huang_clinical_2020}
\bibfield{author}{\bibinfo{person}{Chaolin Huang}, \bibinfo{person}{Yeming
  Wang}, \bibinfo{person}{Xingwang Li}, \bibinfo{person}{Lili Ren},
  \bibinfo{person}{Jianping Zhao}, \bibinfo{person}{Yi Hu}, \bibinfo{person}{Li
  Zhang}, \bibinfo{person}{Guohui Fan}, \bibinfo{person}{Jiuyang Xu}, {and}
  \bibinfo{person}{Xiaoying Gu}.} \bibinfo{year}{2020}\natexlab{}.
\newblock \showarticletitle{Clinical features of patients infected with 2019
  novel coronavirus in {Wuhan}, {China}}.
\newblock \bibinfo{journal}{\emph{The lancet}} \bibinfo{volume}{395},
  \bibinfo{number}{10223} (\bibinfo{year}{2020}), \bibinfo{pages}{497--506}.
\newblock
\newblock
\shownote{ISBN: 0140-6736 Publisher: Elsevier.}


\bibitem[\protect\citeauthoryear{Huang, Starbird, Orand, Stanek, and
  Pedersen}{Huang et~al\mbox{.}}{2015}]%
        {Huang2015}
\bibfield{author}{\bibinfo{person}{Y.~Linlin Huang}, \bibinfo{person}{Kate
  Starbird}, \bibinfo{person}{Mania Orand}, \bibinfo{person}{Stephanie~A.
  Stanek}, {and} \bibinfo{person}{Heather~T. Pedersen}.}
  \bibinfo{year}{2015}\natexlab{}.
\newblock \showarticletitle{Connected Through Crisis: Emotional Proximity and
  the Spread of Misinformation Online}. In
  \bibinfo{booktitle}{\emph{Proceedings of the 18th ACM Conference on Computer
  Supported Cooperative Work \& Social Computing}} (Vancouver, BC, Canada)
  \emph{(\bibinfo{series}{CSCW '15})}. \bibinfo{publisher}{Association for
  Computing Machinery}, \bibinfo{address}{New York, NY, USA},
  \bibinfo{pages}{969–980}.
\newblock
\showISBNx{9781450329224}
\urldef\tempurl%
\url{https://doi.org/10.1145/2675133.2675202}
\showDOI{\tempurl}


\bibitem[\protect\citeauthoryear{Irwin}{Irwin}{1995}]%
        {irwin1995citizen}
\bibfield{author}{\bibinfo{person}{Alan Irwin}.}
  \bibinfo{year}{1995}\natexlab{}.
\newblock \bibinfo{booktitle}{\emph{Citizen science: A study of people,
  expertise and sustainable development}}.
\newblock \bibinfo{publisher}{Psychology Press}.
\newblock


\bibitem[\protect\citeauthoryear{Jackson, \O{}sterlund, Crowston, Harandi, and
  Trouille}{Jackson et~al\mbox{.}}{2020}]%
        {Jackson2020}
\bibfield{author}{\bibinfo{person}{Corey~Brian Jackson},
  \bibinfo{person}{Carsten \O{}sterlund}, \bibinfo{person}{Kevin Crowston},
  \bibinfo{person}{Mahboobeh Harandi}, {and} \bibinfo{person}{Laura Trouille}.}
  \bibinfo{year}{2020}\natexlab{}.
\newblock \showarticletitle{Shifting Forms of Engagement: Volunteer Learning in
  Online Citizen Science}.
\newblock \bibinfo{journal}{\emph{Proc. ACM Hum.-Comput. Interact.}}
  \bibinfo{volume}{4}, \bibinfo{number}{CSCW1}, Article
  \bibinfo{articleno}{036} (\bibinfo{date}{May} \bibinfo{year}{2020}),
  \bibinfo{numpages}{19}~pages.
\newblock
\urldef\tempurl%
\url{https://doi.org/10.1145/3392841}
\showDOI{\tempurl}


\bibitem[\protect\citeauthoryear{Jay, Dunne, Gelsthorpe, and Vigo}{Jay
  et~al\mbox{.}}{2016}]%
        {Jay2016}
\bibfield{author}{\bibinfo{person}{Caroline Jay}, \bibinfo{person}{Robert
  Dunne}, \bibinfo{person}{David Gelsthorpe}, {and} \bibinfo{person}{Markel
  Vigo}.} \bibinfo{year}{2016}\natexlab{}.
\newblock \showarticletitle{To Sign Up, or Not to Sign Up? Maximizing Citizen
  Science Contribution Rates through Optional Registration}. In
  \bibinfo{booktitle}{\emph{Proceedings of the 2016 CHI Conference on Human
  Factors in Computing Systems}} (San Jose, California, USA)
  \emph{(\bibinfo{series}{CHI '16})}. \bibinfo{publisher}{Association for
  Computing Machinery}, \bibinfo{address}{New York, NY, USA},
  \bibinfo{pages}{1827–1832}.
\newblock
\showISBNx{9781450333627}
\urldef\tempurl%
\url{https://doi.org/10.1145/2858036.2858319}
\showDOI{\tempurl}


\bibitem[\protect\citeauthoryear{Jennett, Kloetzer, Schneider, Iacovides, Cox,
  Gold, Fuchs, Eveleigh, Methieu, Ajani, et~al\mbox{.}}{Jennett
  et~al\mbox{.}}{2016}]%
        {jennett2016motivations}
\bibfield{author}{\bibinfo{person}{Charlene Jennett}, \bibinfo{person}{Laure
  Kloetzer}, \bibinfo{person}{Daniel Schneider}, \bibinfo{person}{Ioanna
  Iacovides}, \bibinfo{person}{Anna Cox}, \bibinfo{person}{Margaret Gold},
  \bibinfo{person}{Brian Fuchs}, \bibinfo{person}{Alexandra Eveleigh},
  \bibinfo{person}{Kathleen Methieu}, \bibinfo{person}{Zoya Ajani},
  {et~al\mbox{.}}} \bibinfo{year}{2016}\natexlab{}.
\newblock \showarticletitle{Motivations, learning and creativity in online
  citizen science}.
\newblock \bibinfo{journal}{\emph{Journal of Science Communication}}
  \bibinfo{volume}{15}, \bibinfo{number}{3} (\bibinfo{year}{2016}).
\newblock


\bibitem[\protect\citeauthoryear{{Johns Hopkins University}}{{Johns Hopkins
  University}}{2020}]%
        {johns2020covid}
\bibfield{author}{\bibinfo{person}{{Johns Hopkins University}}.}
  \bibinfo{year}{2020}\natexlab{}.
\newblock \showarticletitle{COVID-19 Dashboard by the Center for Systems
  Science and Engineering (CSSE) at Johns Hopkins University (JHU)}.
\newblock  (\bibinfo{year}{2020}).
\newblock


\bibitem[\protect\citeauthoryear{Johnson, McMahon, Sch{\"o}ning, and
  Hecht}{Johnson et~al\mbox{.}}{2017}]%
        {johnson2017effect}
\bibfield{author}{\bibinfo{person}{Isaac Johnson}, \bibinfo{person}{Connor
  McMahon}, \bibinfo{person}{Johannes Sch{\"o}ning}, {and}
  \bibinfo{person}{Brent Hecht}.} \bibinfo{year}{2017}\natexlab{}.
\newblock \showarticletitle{The Effect of Population and" Structural" Biases on
  Social Media-based Algorithms: A Case Study in Geolocation Inference Across
  the Urban-Rural Spectrum}. In \bibinfo{booktitle}{\emph{Proceedings of the
  2017 CHI conference on Human Factors in Computing Systems}}.
  \bibinfo{pages}{1167--1178}.
\newblock


\bibitem[\protect\citeauthoryear{Johnson, Lin, Li, Hall, Halfaker,
  Sch{\"o}ning, and Hecht}{Johnson et~al\mbox{.}}{2016}]%
        {johnson2016not}
\bibfield{author}{\bibinfo{person}{Isaac~L Johnson}, \bibinfo{person}{Yilun
  Lin}, \bibinfo{person}{Toby Jia-Jun Li}, \bibinfo{person}{Andrew Hall},
  \bibinfo{person}{Aaron Halfaker}, \bibinfo{person}{Johannes Sch{\"o}ning},
  {and} \bibinfo{person}{Brent Hecht}.} \bibinfo{year}{2016}\natexlab{}.
\newblock \showarticletitle{Not at home on the range: Peer production and the
  urban/rural divide}. In \bibinfo{booktitle}{\emph{Proceedings of the 2016 CHI
  Conference on Human Factors in Computing Systems}}. \bibinfo{pages}{13--25}.
\newblock


\bibitem[\protect\citeauthoryear{Karapanos, Zimmerman, Forlizzi, and
  Martens}{Karapanos et~al\mbox{.}}{2009}]%
        {Karapanos2009}
\bibfield{author}{\bibinfo{person}{Evangelos Karapanos}, \bibinfo{person}{John
  Zimmerman}, \bibinfo{person}{Jodi Forlizzi}, {and}
  \bibinfo{person}{Jean-Bernard Martens}.} \bibinfo{year}{2009}\natexlab{}.
\newblock \showarticletitle{User Experience over Time: An Initial Framework}.
  In \bibinfo{booktitle}{\emph{Proceedings of the SIGCHI Conference on Human
  Factors in Computing Systems}} (Boston, MA, USA) \emph{(\bibinfo{series}{CHI
  '09})}. \bibinfo{publisher}{Association for Computing Machinery},
  \bibinfo{address}{New York, NY, USA}, \bibinfo{pages}{729–738}.
\newblock
\showISBNx{9781605582467}
\urldef\tempurl%
\url{https://doi.org/10.1145/1518701.1518814}
\showDOI{\tempurl}


\bibitem[\protect\citeauthoryear{Kim, Robson, Zimmerman, Pierce, and Haber}{Kim
  et~al\mbox{.}}{2011}]%
        {kim2011creek}
\bibfield{author}{\bibinfo{person}{Sunyoung Kim}, \bibinfo{person}{Christine
  Robson}, \bibinfo{person}{Thomas Zimmerman}, \bibinfo{person}{Jeffrey
  Pierce}, {and} \bibinfo{person}{Eben~M Haber}.}
  \bibinfo{year}{2011}\natexlab{}.
\newblock \showarticletitle{Creek watch: pairing usefulness and usability for
  successful citizen science}. In \bibinfo{booktitle}{\emph{Proceedings of the
  SIGCHI Conference on Human Factors in Computing Systems}}.
  \bibinfo{pages}{2125--2134}.
\newblock


\bibitem[\protect\citeauthoryear{Klandermans}{Klandermans}{1997}]%
        {klandermans1997social}
\bibfield{author}{\bibinfo{person}{Bert Klandermans}.}
  \bibinfo{year}{1997}\natexlab{}.
\newblock \bibinfo{booktitle}{\emph{The social psychology of protest}}.
\newblock \bibinfo{publisher}{Blackwell}, \bibinfo{address}{Oxford, UK}.
\newblock


\bibitem[\protect\citeauthoryear{Land-Zandstra, Devilee, Snik, Buurmeijer, and
  van~den Broek}{Land-Zandstra et~al\mbox{.}}{2016}]%
        {land2016citizen}
\bibfield{author}{\bibinfo{person}{Anne~M Land-Zandstra},
  \bibinfo{person}{Jeroen~LA Devilee}, \bibinfo{person}{Frans Snik},
  \bibinfo{person}{Franka Buurmeijer}, {and} \bibinfo{person}{Jos~M van~den
  Broek}.} \bibinfo{year}{2016}\natexlab{}.
\newblock \showarticletitle{Citizen science on a smartphone: Participants’
  motivations and learning}.
\newblock \bibinfo{journal}{\emph{Public Understanding of Science}}
  \bibinfo{volume}{25}, \bibinfo{number}{1} (\bibinfo{year}{2016}),
  \bibinfo{pages}{45--60}.
\newblock


\bibitem[\protect\citeauthoryear{MDR}{MDR}{2020}]%
        {MDR2020chronik}
\bibfield{author}{\bibinfo{person}{MDR}.} \bibinfo{year}{2020}\natexlab{}.
\newblock \bibinfo{booktitle}{\emph{Die Chronik der Corona-Krise}}.
\newblock
\urldef\tempurl%
\url{https://www.mdr.de/nachrichten/politik/corona-chronik-chronologie-coronavirus-100.html}
\showURL{%
Retrieved September 14, 2020 from \tempurl}


\bibitem[\protect\citeauthoryear{Nov}{Nov}{2007}]%
        {nov2007motivates}
\bibfield{author}{\bibinfo{person}{Oded Nov}.} \bibinfo{year}{2007}\natexlab{}.
\newblock \showarticletitle{What motivates wikipedians?}
\newblock \bibinfo{journal}{\emph{Commun. ACM}} \bibinfo{volume}{50},
  \bibinfo{number}{11} (\bibinfo{year}{2007}), \bibinfo{pages}{60--64}.
\newblock


\bibitem[\protect\citeauthoryear{Nov, Arazy, and Anderson}{Nov
  et~al\mbox{.}}{2011}]%
        {nov2011dusting}
\bibfield{author}{\bibinfo{person}{Oded Nov}, \bibinfo{person}{Ofer Arazy},
  {and} \bibinfo{person}{David Anderson}.} \bibinfo{year}{2011}\natexlab{}.
\newblock \showarticletitle{Dusting for science: motivation and participation
  of digital citizen science volunteers}.
\newblock In \bibinfo{booktitle}{\emph{Proceedings of the 2011 iConference}}.
  \bibinfo{pages}{68--74}.
\newblock


\bibitem[\protect\citeauthoryear{of~Life}{of~Life}{2020}]%
        {encyclopediaOfLife}
\bibfield{author}{\bibinfo{person}{Encyclopedia of Life}.}
  \bibinfo{year}{2020}\natexlab{}.
\newblock \bibinfo{booktitle}{\emph{Encyclopedia of Life}}.
\newblock
\urldef\tempurl%
\url{https://eol.org/}
\showURL{%
Retrieved September 14, 2020 from \tempurl}


\bibitem[\protect\citeauthoryear{of~Ornithology}{of~Ornithology}{2020}]%
        {eBird}
\bibfield{author}{\bibinfo{person}{The Cornell~Lab of Ornithology}.}
  \bibinfo{year}{2020}\natexlab{}.
\newblock \bibinfo{booktitle}{\emph{eBird - Discover a new world of
  birding...}}
\newblock
\urldef\tempurl%
\url{https://ebird.org/home}
\showURL{%
Retrieved September 14, 2020 from \tempurl}


\bibitem[\protect\citeauthoryear{of~the Interior}{of~the Interior}{2020}]%
        {nationalparksservice}
\bibfield{author}{\bibinfo{person}{U.S.~Department of~the Interior}.}
  \bibinfo{year}{2020}\natexlab{}.
\newblock \bibinfo{booktitle}{\emph{Gulf Coast Inventory \& Monitoring Network
  (U.S. National Park Service)}}.
\newblock
\urldef\tempurl%
\url{https://www.nps.gov/im/guln/index.htm}
\showURL{%
Retrieved September 14, 2020 from \tempurl}


\bibitem[\protect\citeauthoryear{Organization}{Organization}{2019}]%
        {WHOmeasles}
\bibfield{author}{\bibinfo{person}{World~Health Organization}.}
  \bibinfo{year}{2019}\natexlab{}.
\newblock \bibinfo{booktitle}{\emph{Measles}}.
\newblock
\urldef\tempurl%
\url{https://www.who.int/news-room/fact-sheets/detail/measles}
\showURL{%
Retrieved December 28, 2020 from \tempurl}


\bibitem[\protect\citeauthoryear{Preece}{Preece}{2016}]%
        {preece2016citizen}
\bibfield{author}{\bibinfo{person}{Jennifer Preece}.}
  \bibinfo{year}{2016}\natexlab{}.
\newblock \showarticletitle{Citizen science: New research challenges for
  human--computer interaction}.
\newblock \bibinfo{journal}{\emph{International Journal of Human-Computer
  Interaction}} \bibinfo{volume}{32}, \bibinfo{number}{8}
  (\bibinfo{year}{2016}), \bibinfo{pages}{585--612}.
\newblock


\bibitem[\protect\citeauthoryear{Qkopy}{Qkopy}{2020}]%
        {GoKDirect}
\bibfield{author}{\bibinfo{person}{Qkopy}.} \bibinfo{year}{2020}\natexlab{}.
\newblock \bibinfo{booktitle}{\emph{GoK Direct - Kerala - Apps on Google
  Play}}.
\newblock
\urldef\tempurl%
\url{https://play.google.com/store/apps/details?id=com.qkopy.prdkerala&hl=en}
\showURL{%
Retrieved September 14, 2020 from \tempurl}


\bibitem[\protect\citeauthoryear{Quer, Radin, Gadaleta, Baca-Motes, Ariniello,
  Ramos, Kheterpal, Topol, and Steinhubl}{Quer et~al\mbox{.}}{2021}]%
        {quer2020passive}
\bibfield{author}{\bibinfo{person}{Giorgio Quer}, \bibinfo{person}{Jennifer~M.
  Radin}, \bibinfo{person}{Matteo Gadaleta}, \bibinfo{person}{Katie
  Baca-Motes}, \bibinfo{person}{Lauren Ariniello}, \bibinfo{person}{Edward
  Ramos}, \bibinfo{person}{Vik Kheterpal}, \bibinfo{person}{Eric~J. Topol},
  {and} \bibinfo{person}{Steven~R. Steinhubl}.}
  \bibinfo{year}{2021}\natexlab{}.
\newblock \showarticletitle{Wearable sensor data and self-reported symptoms for
  {COVID}-19 detection}.
\newblock \bibinfo{journal}{\emph{Nature Medicine}} \bibinfo{volume}{27},
  \bibinfo{number}{1} (\bibinfo{year}{2021}), \bibinfo{pages}{73--77}.
\newblock
\showISSN{1546-170X}
\urldef\tempurl%
\url{https://doi.org/10.1038/s41591-020-1123-x}
\showDOI{\tempurl}


\bibitem[\protect\citeauthoryear{Raddick, Bracey, Gay, Lintott, Cardamone,
  Murray, Schawinski, Szalay, and Vandenberg}{Raddick et~al\mbox{.}}{2013}]%
        {raddick2013galaxy}
\bibfield{author}{\bibinfo{person}{Jordan Raddick}, \bibinfo{person}{Georgia
  Bracey}, \bibinfo{person}{Pamela Gay}, \bibinfo{person}{Chris Lintott},
  \bibinfo{person}{Carolin Cardamone}, \bibinfo{person}{Phil Murray},
  \bibinfo{person}{Kevin Schawinski}, \bibinfo{person}{Alexander Szalay}, {and}
  \bibinfo{person}{Jan Vandenberg}.} \bibinfo{year}{2013}\natexlab{}.
\newblock \showarticletitle{Galaxy Zoo: Motivations of Citizen Scientists}.
\newblock \bibinfo{journal}{\emph{Astronomy Education Review}}
  \bibinfo{volume}{12} (\bibinfo{date}{12} \bibinfo{year}{2013}),
  \bibinfo{pages}{0106--}.
\newblock
\urldef\tempurl%
\url{https://doi.org/10.3847/AER2011021}
\showDOI{\tempurl}


\bibitem[\protect\citeauthoryear{Radin, Wineinger, Topol, and Steinhubl}{Radin
  et~al\mbox{.}}{2020}]%
        {radin2020harnessing}
\bibfield{author}{\bibinfo{person}{Jennifer~M Radin}, \bibinfo{person}{Nathan~E
  Wineinger}, \bibinfo{person}{Eric~J Topol}, {and} \bibinfo{person}{Steven~R
  Steinhubl}.} \bibinfo{year}{2020}\natexlab{}.
\newblock \showarticletitle{Harnessing wearable device data to improve
  state-level real-time surveillance of influenza-like illness in the USA: a
  population-based study}.
\newblock \bibinfo{journal}{\emph{The Lancet Digital Health}}
  \bibinfo{volume}{2}, \bibinfo{number}{2} (\bibinfo{year}{2020}),
  \bibinfo{pages}{e85--e93}.
\newblock


\bibitem[\protect\citeauthoryear{Reeves and Simperl}{Reeves and
  Simperl}{2019}]%
        {Reeves2019}
\bibfield{author}{\bibinfo{person}{Neal~T. Reeves} {and} \bibinfo{person}{Elena
  Simperl}.} \bibinfo{year}{2019}\natexlab{}.
\newblock \showarticletitle{Efficient, but Effective? Volunteer Engagement in
  Short-Term Virtual Citizen Science Projects}.
\newblock \bibinfo{journal}{\emph{Proc. ACM Hum.-Comput. Interact.}}
  \bibinfo{volume}{3}, \bibinfo{number}{CSCW}, Article \bibinfo{articleno}{177}
  (\bibinfo{date}{Nov.} \bibinfo{year}{2019}), \bibinfo{numpages}{35}~pages.
\newblock
\urldef\tempurl%
\url{https://doi.org/10.1145/3359279}
\showDOI{\tempurl}


\bibitem[\protect\citeauthoryear{RKI}{RKI}{2020a}]%
        {feverCurve}
\bibfield{author}{\bibinfo{person}{RKI}.} \bibinfo{year}{2020}\natexlab{a}.
\newblock \bibinfo{booktitle}{\emph{Aktuelle Fieberkurve für Deutschland}}.
\newblock
\urldef\tempurl%
\url{https://corona-datenspende.de/science/monitor/}
\showURL{%
Retrieved September 14, 2020 from \tempurl}


\bibitem[\protect\citeauthoryear{RKI}{RKI}{2020b}]%
        {coronawarnapp}
\bibfield{author}{\bibinfo{person}{RKI}.} \bibinfo{year}{2020}\natexlab{b}.
\newblock \bibinfo{booktitle}{\emph{Corona-Warn-App}}.
\newblock
\urldef\tempurl%
\url{https://www.rki.de/DE/Content/InfAZ/N/Neuartiges_Coronavirus/WarnApp/Warn_App.html}
\showURL{%
Retrieved September 14, 2020 from \tempurl}


\bibitem[\protect\citeauthoryear{RKI}{RKI}{2020c}]%
        {coronadatadonationapp}
\bibfield{author}{\bibinfo{person}{RKI}.} \bibinfo{year}{2020}\natexlab{c}.
\newblock \bibinfo{booktitle}{\emph{Datenspende: Robert Koch-Institut -
  Corona-Datenspende}}.
\newblock
\urldef\tempurl%
\url{https://corona-datenspende.de/}
\showURL{%
Retrieved September 14, 2020 from \tempurl}


\bibitem[\protect\citeauthoryear{RKI}{RKI}{2020d}]%
        {rkiThankYou}
\bibfield{author}{\bibinfo{person}{RKI}.} \bibinfo{year}{2020}\natexlab{d}.
\newblock \bibinfo{booktitle}{\emph{Ein Dankeschön an die Spender:innen!}}
\newblock
\urldef\tempurl%
\url{https://corona-datenspende.de/science/reports/thank-you}
\showURL{%
Retrieved September 14, 2020 from \tempurl}


\bibitem[\protect\citeauthoryear{RKI}{RKI}{2020e}]%
        {rkiDatenspendeApp}
\bibfield{author}{\bibinfo{person}{RKI}.} \bibinfo{year}{2020}\natexlab{e}.
\newblock \bibinfo{booktitle}{\emph{Robert Koch Institute -
  Corona-Datenspende-App}}.
\newblock
\urldef\tempurl%
\url{https://www.rki.de/DE/Content/InfAZ/N/Neuartiges_Coronavirus/Corona-Datenspende-allgemein.html}
\showURL{%
Retrieved September 14, 2020 from \tempurl}


\bibitem[\protect\citeauthoryear{Roberts, Hann, and Slaughter}{Roberts
  et~al\mbox{.}}{2006}]%
        {roberts2006understanding}
\bibfield{author}{\bibinfo{person}{Jeffrey~A Roberts}, \bibinfo{person}{Il-Horn
  Hann}, {and} \bibinfo{person}{Sandra~A Slaughter}.}
  \bibinfo{year}{2006}\natexlab{}.
\newblock \showarticletitle{Understanding the motivations, participation, and
  performance of open source software developers: A longitudinal study of the
  Apache projects}.
\newblock \bibinfo{journal}{\emph{Management science}} \bibinfo{volume}{52},
  \bibinfo{number}{7} (\bibinfo{year}{2006}), \bibinfo{pages}{984--999}.
\newblock


\bibitem[\protect\citeauthoryear{Rotman, Hammock, Preece, Hansen, Boston,
  Bowser, and He}{Rotman et~al\mbox{.}}{2014}]%
        {rotman2014motivations}
\bibfield{author}{\bibinfo{person}{Dana Rotman}, \bibinfo{person}{Jen Hammock},
  \bibinfo{person}{Jenny Preece}, \bibinfo{person}{Derek Hansen},
  \bibinfo{person}{Carol Boston}, \bibinfo{person}{Anne Bowser}, {and}
  \bibinfo{person}{Yurong He}.} \bibinfo{year}{2014}\natexlab{}.
\newblock \showarticletitle{Motivations affecting initial and long-term
  participation in citizen science projects in three countries}.
\newblock \bibinfo{journal}{\emph{IConference 2014 Proceedings}}
  (\bibinfo{year}{2014}).
\newblock


\bibitem[\protect\citeauthoryear{Rotman, Preece, Hammock, Procita, Hansen,
  Parr, Lewis, and Jacobs}{Rotman et~al\mbox{.}}{2012}]%
        {rotman2012dynamic}
\bibfield{author}{\bibinfo{person}{Dana Rotman}, \bibinfo{person}{Jenny
  Preece}, \bibinfo{person}{Jen Hammock}, \bibinfo{person}{Kezee Procita},
  \bibinfo{person}{Derek Hansen}, \bibinfo{person}{Cynthia Parr},
  \bibinfo{person}{Darcy Lewis}, {and} \bibinfo{person}{David Jacobs}.}
  \bibinfo{year}{2012}\natexlab{}.
\newblock \showarticletitle{Dynamic changes in motivation in collaborative
  citizen-science projects}. In \bibinfo{booktitle}{\emph{Proceedings of the
  ACM 2012 conference on computer supported cooperative work}}.
  \bibinfo{pages}{217--226}.
\newblock


\bibitem[\protect\citeauthoryear{Schroer and Hertel}{Schroer and
  Hertel}{2009}]%
        {schroer2009voluntary}
\bibfield{author}{\bibinfo{person}{Joachim Schroer} {and}
  \bibinfo{person}{Guido Hertel}.} \bibinfo{year}{2009}\natexlab{}.
\newblock \showarticletitle{Voluntary engagement in an open web-based
  encyclopedia: Wikipedians and why they do it}.
\newblock \bibinfo{journal}{\emph{Media Psychology}} \bibinfo{volume}{12},
  \bibinfo{number}{1} (\bibinfo{year}{2009}), \bibinfo{pages}{96--120}.
\newblock


\bibitem[\protect\citeauthoryear{Simperl, Reeves, Phethean, Lynes, and
  Tinati}{Simperl et~al\mbox{.}}{2018}]%
        {SimperlGame2018}
\bibfield{author}{\bibinfo{person}{Elena Simperl}, \bibinfo{person}{Neal
  Reeves}, \bibinfo{person}{Chris Phethean}, \bibinfo{person}{Todd Lynes},
  {and} \bibinfo{person}{Ramine Tinati}.} \bibinfo{year}{2018}\natexlab{}.
\newblock \showarticletitle{Is Virtual Citizen Science A Game?}
\newblock \bibinfo{journal}{\emph{Trans. Soc. Comput.}} \bibinfo{volume}{1},
  \bibinfo{number}{2}, Article \bibinfo{articleno}{6} (\bibinfo{date}{June}
  \bibinfo{year}{2018}), \bibinfo{numpages}{39}~pages.
\newblock
\showISSN{2469-7818}
\urldef\tempurl%
\url{https://doi.org/10.1145/3209960}
\showDOI{\tempurl}


\bibitem[\protect\citeauthoryear{Snyder}{Snyder}{2017}]%
        {Snyder2017}
\bibfield{author}{\bibinfo{person}{Jaime Snyder}.}
  \bibinfo{year}{2017}\natexlab{}.
\newblock \showarticletitle{Vernacular Visualization Practices in a Citizen
  Science Project}. In \bibinfo{booktitle}{\emph{Proceedings of the 2017 ACM
  Conference on Computer Supported Cooperative Work and Social Computing}}
  (Portland, Oregon, USA) \emph{(\bibinfo{series}{CSCW '17})}.
  \bibinfo{publisher}{Association for Computing Machinery},
  \bibinfo{address}{New York, NY, USA}, \bibinfo{pages}{2097–2111}.
\newblock
\showISBNx{9781450343350}
\urldef\tempurl%
\url{https://doi.org/10.1145/2998181.2998239}
\showDOI{\tempurl}


\bibitem[\protect\citeauthoryear{Soden, Hamel, Lallemant, and Pierce}{Soden
  et~al\mbox{.}}{2020}]%
        {Soden-et-al-2020}
\bibfield{author}{\bibinfo{person}{Robert Soden}, \bibinfo{person}{Perrine
  Hamel}, \bibinfo{person}{David Lallemant}, {and} \bibinfo{person}{James
  Pierce}.} \bibinfo{year}{2020}\natexlab{}.
\newblock \showarticletitle{The Disaster and Climate Change Artathon: Staging
  Art/Science Collaborations in Crisis Informatics}. In
  \bibinfo{booktitle}{\emph{Proceedings of the 2020 ACM Designing Interactive
  Systems Conference}} (Eindhoven, Netherlands) \emph{(\bibinfo{series}{DIS
  '20})}. \bibinfo{publisher}{Association for Computing Machinery},
  \bibinfo{address}{New York, NY, USA}, \bibinfo{pages}{1273–1286}.
\newblock
\showISBNx{9781450369749}
\urldef\tempurl%
\url{https://doi.org/10.1145/3357236.3395461}
\showDOI{\tempurl}


\bibitem[\protect\citeauthoryear{Soden and Palen}{Soden and Palen}{2018}]%
        {Soden2018}
\bibfield{author}{\bibinfo{person}{Robert Soden} {and} \bibinfo{person}{Leysia
  Palen}.} \bibinfo{year}{2018}\natexlab{}.
\newblock \showarticletitle{Informating Crisis: Expanding Critical Perspectives
  in Crisis Informatics}.
\newblock \bibinfo{journal}{\emph{Proc. ACM Hum.-Comput. Interact.}}
  \bibinfo{volume}{2}, \bibinfo{number}{CSCW}, Article \bibinfo{articleno}{162}
  (\bibinfo{date}{Nov.} \bibinfo{year}{2018}), \bibinfo{numpages}{22}~pages.
\newblock
\urldef\tempurl%
\url{https://doi.org/10.1145/3274431}
\showDOI{\tempurl}


\bibitem[\protect\citeauthoryear{Starbird and Palen}{Starbird and
  Palen}{2011}]%
        {starbird2011voluntweeters}
\bibfield{author}{\bibinfo{person}{Kate Starbird} {and} \bibinfo{person}{Leysia
  Palen}.} \bibinfo{year}{2011}\natexlab{}.
\newblock \showarticletitle{"Voluntweeters": Self-Organizing by Digital
  Volunteers in Times of Crisis}. In \bibinfo{booktitle}{\emph{Proceedings of
  the SIGCHI Conference on Human Factors in Computing Systems}} (Vancouver, BC,
  Canada) \emph{(\bibinfo{series}{CHI '11})}. \bibinfo{publisher}{Association
  for Computing Machinery}, \bibinfo{address}{New York, NY, USA},
  \bibinfo{pages}{1071–1080}.
\newblock
\showISBNx{9781450302289}
\urldef\tempurl%
\url{https://doi.org/10.1145/1978942.1979102}
\showDOI{\tempurl}


\bibitem[\protect\citeauthoryear{Stawarz, Cox, and Blandford}{Stawarz
  et~al\mbox{.}}{2014}]%
        {Stawarz2014pill}
\bibfield{author}{\bibinfo{person}{Katarzyna Stawarz}, \bibinfo{person}{Anna~L.
  Cox}, {and} \bibinfo{person}{Ann Blandford}.}
  \bibinfo{year}{2014}\natexlab{}.
\newblock \showarticletitle{Don't Forget Your Pill! Designing Effective
  Medication Reminder Apps That Support Users' Daily Routines}. In
  \bibinfo{booktitle}{\emph{Proceedings of the SIGCHI Conference on Human
  Factors in Computing Systems}} (Toronto, Ontario, Canada)
  \emph{(\bibinfo{series}{CHI '14})}. \bibinfo{publisher}{Association for
  Computing Machinery}, \bibinfo{address}{New York, NY, USA},
  \bibinfo{pages}{2269–2278}.
\newblock
\showISBNx{9781450324731}
\urldef\tempurl%
\url{https://doi.org/10.1145/2556288.2557079}
\showDOI{\tempurl}


\bibitem[\protect\citeauthoryear{Sullivan, Wood, Iliff, Bonney, Fink, and
  Kelling}{Sullivan et~al\mbox{.}}{2009}]%
        {sullivan2009ebird}
\bibfield{author}{\bibinfo{person}{Brian~L Sullivan},
  \bibinfo{person}{Christopher~L Wood}, \bibinfo{person}{Marshall~J Iliff},
  \bibinfo{person}{Rick~E Bonney}, \bibinfo{person}{Daniel Fink}, {and}
  \bibinfo{person}{Steve Kelling}.} \bibinfo{year}{2009}\natexlab{}.
\newblock \showarticletitle{eBird: A citizen-based bird observation network in
  the biological sciences}.
\newblock \bibinfo{journal}{\emph{Biological conservation}}
  \bibinfo{volume}{142}, \bibinfo{number}{10} (\bibinfo{year}{2009}),
  \bibinfo{pages}{2282--2292}.
\newblock


\bibitem[\protect\citeauthoryear{{tagesschau.de}}{{tagesschau.de}}{2020a}]%
        {tagesschau2020datenspende}
\bibfield{author}{\bibinfo{person}{{tagesschau.de}}.}
  \bibinfo{year}{2020}\natexlab{a}.
\newblock \bibinfo{booktitle}{\emph{Hunderttausende nutzen "Datenspende"-App}}.
\newblock
\urldef\tempurl%
\url{https://www.tagesschau.de/inland/datenspende-app-rki-corona-101.html}
\showURL{%
Retrieved September 14, 2020 from \tempurl}


\bibitem[\protect\citeauthoryear{{tagesschau.de}}{{tagesschau.de}}{2020b}]%
        {tagesschau2020datenspendeOnline}
\bibfield{author}{\bibinfo{person}{{tagesschau.de}}.}
  \bibinfo{year}{2020}\natexlab{b}.
\newblock \bibinfo{booktitle}{\emph{Mit Fitnessdaten das Virus verstehen}}.
\newblock
\urldef\tempurl%
\url{https://www.tagesschau.de/inland/app-rki-101.html}
\showURL{%
Retrieved December 17, 2020 from \tempurl}


\bibitem[\protect\citeauthoryear{{tagesschau.de}}{{tagesschau.de}}{2020c}]%
        {tagesschau2020datenspendeTV}
\bibfield{author}{\bibinfo{person}{{tagesschau.de}}.}
  \bibinfo{year}{2020}\natexlab{c}.
\newblock \bibinfo{booktitle}{\emph{Sendung: tagesschau 07.04.2020 20:00 Uhr}}.
\newblock
\urldef\tempurl%
\url{https://www.tagesschau.de/multimedia/sendung/ts-36515.html}
\showURL{%
Retrieved December 17, 2020 from \tempurl}


\bibitem[\protect\citeauthoryear{Tan, Prasanna, Stock, Hudson-Doyle, Leonard,
  and Johnston}{Tan et~al\mbox{.}}{2017}]%
        {tan2017mobile}
\bibfield{author}{\bibinfo{person}{Marion~Lara Tan}, \bibinfo{person}{Raj
  Prasanna}, \bibinfo{person}{Kristin Stock}, \bibinfo{person}{Emma
  Hudson-Doyle}, \bibinfo{person}{Graham Leonard}, {and} \bibinfo{person}{David
  Johnston}.} \bibinfo{year}{2017}\natexlab{}.
\newblock \showarticletitle{Mobile applications in crisis informatics
  literature: A systematic review}.
\newblock \bibinfo{journal}{\emph{International journal of disaster risk
  reduction}}  \bibinfo{volume}{24} (\bibinfo{year}{2017}),
  \bibinfo{pages}{297--311}.
\newblock


\bibitem[\protect\citeauthoryear{Times}{Times}{2020}]%
        {pti_2020}
\bibfield{author}{\bibinfo{person}{Economic Times}.}
  \bibinfo{year}{2020}\natexlab{}.
\newblock \bibinfo{booktitle}{\emph{Those in home quarantine in Karnataka
  directed to send selfies every hour to govt}}.
\newblock
\urldef\tempurl%
\url{https://economictimes.indiatimes.com/news/politics-and-nation/those-in-home-quarantine-in-karnataka-directed-to-send-selfies-every-hour-to-govt/articleshow/74907051.cms?from=mdr}
\showURL{%
Retrieved September 14, 2020 from \tempurl}


\bibitem[\protect\citeauthoryear{Vaast and Urquhart}{Vaast and
  Urquhart}{2017}]%
        {vaast2017building}
\bibfield{author}{\bibinfo{person}{Emmanuelle Vaast} {and}
  \bibinfo{person}{Cathy Urquhart}.} \bibinfo{year}{2017}\natexlab{}.
\newblock \showarticletitle{Building Grounded Theory with Social Media Data}.
\newblock In \bibinfo{booktitle}{\emph{Routledge Companion to Qualitative
  Research in Organization Studies}}, \bibfield{editor}{\bibinfo{person}{Raza
  Mir} {and} \bibinfo{person}{Sanjay Jain}} (Eds.).
  \bibinfo{publisher}{Routledge}, Chapter~14.
\newblock


\bibitem[\protect\citeauthoryear{Vaishya, Javaid, Khan, and Haleem}{Vaishya
  et~al\mbox{.}}{2020}]%
        {vaishya2020artificial}
\bibfield{author}{\bibinfo{person}{Raju Vaishya}, \bibinfo{person}{Mohd
  Javaid}, \bibinfo{person}{Ibrahim~Haleem Khan}, {and} \bibinfo{person}{Abid
  Haleem}.} \bibinfo{year}{2020}\natexlab{}.
\newblock \showarticletitle{Artificial Intelligence (AI) applications for
  COVID-19 pandemic}.
\newblock \bibinfo{journal}{\emph{Diabetes \& Metabolic Syndrome: Clinical
  Research \& Reviews}} \bibinfo{volume}{14}, \bibinfo{number}{4}
  (\bibinfo{year}{2020}).
\newblock


\bibitem[\protect\citeauthoryear{Vieweg, Hughes, Starbird, and Palen}{Vieweg
  et~al\mbox{.}}{2010}]%
        {Vieweg2010}
\bibfield{author}{\bibinfo{person}{Sarah Vieweg}, \bibinfo{person}{Amanda~L.
  Hughes}, \bibinfo{person}{Kate Starbird}, {and} \bibinfo{person}{Leysia
  Palen}.} \bibinfo{year}{2010}\natexlab{}.
\newblock \showarticletitle{Microblogging during Two Natural Hazards Events:
  What Twitter May Contribute to Situational Awareness}. In
  \bibinfo{booktitle}{\emph{Proceedings of the SIGCHI Conference on Human
  Factors in Computing Systems}} (Atlanta, Georgia, USA)
  \emph{(\bibinfo{series}{CHI '10})}. \bibinfo{publisher}{Association for
  Computing Machinery}, \bibinfo{address}{New York, NY, USA},
  \bibinfo{pages}{1079–1088}.
\newblock
\showISBNx{9781605589299}
\urldef\tempurl%
\url{https://doi.org/10.1145/1753326.1753486}
\showDOI{\tempurl}


\bibitem[\protect\citeauthoryear{Whitelaw, Mamas, Topol, and
  Van~Spall}{Whitelaw et~al\mbox{.}}{2020}]%
        {whitelaw2020applications}
\bibfield{author}{\bibinfo{person}{Sera Whitelaw}, \bibinfo{person}{Mamas~A
  Mamas}, \bibinfo{person}{Eric Topol}, {and} \bibinfo{person}{Harriette~GC
  Van~Spall}.} \bibinfo{year}{2020}\natexlab{}.
\newblock \showarticletitle{Applications of digital technology in COVID-19
  pandemic planning and response}.
\newblock \bibinfo{journal}{\emph{The Lancet Digital Health}}
  (\bibinfo{year}{2020}).
\newblock


\bibitem[\protect\citeauthoryear{WHO}{WHO}{2020a}]%
        {WHONovelCovid}
\bibfield{author}{\bibinfo{person}{WHO}.} \bibinfo{year}{2020}\natexlab{a}.
\newblock \bibinfo{booktitle}{\emph{Novel Coronavirus – China}}.
\newblock
\urldef\tempurl%
\url{http://www.who.int/csr/don/12-january-2020-novel-coronavirus-china/en/}
\showURL{%
Retrieved September 14, 2020 from \tempurl}


\bibitem[\protect\citeauthoryear{WHO}{WHO}{2020b}]%
        {WHOCovidTransmitted}
\bibfield{author}{\bibinfo{person}{WHO}.} \bibinfo{year}{2020}\natexlab{b}.
\newblock \bibinfo{booktitle}{\emph{Q\&A: How is COVID-19 transmitted?}}
\newblock
\urldef\tempurl%
\url{https://www.who.int/news-room/q-a-detail/q-a-how-is-covid-19-transmitted}
\showURL{%
Retrieved September 14, 2020 from \tempurl}


\bibitem[\protect\citeauthoryear{WHO}{WHO}{2020c}]%
        {WHOQACorona}
\bibfield{author}{\bibinfo{person}{WHO}.} \bibinfo{year}{2020}\natexlab{c}.
\newblock \bibinfo{booktitle}{\emph{Q\&A on coronaviruses (COVID-19)}}.
\newblock
\urldef\tempurl%
\url{https://www.who.int/news-room/q-a-detail/q-a-coronaviruses}
\showURL{%
Retrieved September 14, 2020 from \tempurl}


\bibitem[\protect\citeauthoryear{WHO}{WHO}{2020d}]%
        {worldhealthorganizationdashboard}
\bibfield{author}{\bibinfo{person}{WHO}.} \bibinfo{year}{2020}\natexlab{d}.
\newblock \bibinfo{booktitle}{\emph{WHO Coronavirus Disease (COVID-19)
  Dashboard}}.
\newblock
\urldef\tempurl%
\url{hhttps://covid19.who.int/}
\showURL{%
Retrieved September 14, 2020 from \tempurl}


\bibitem[\protect\citeauthoryear{Wiggins and Crowston}{Wiggins and
  Crowston}{2011}]%
        {wiggins2011conservation}
\bibfield{author}{\bibinfo{person}{Andrea Wiggins} {and} \bibinfo{person}{Kevin
  Crowston}.} \bibinfo{year}{2011}\natexlab{}.
\newblock \showarticletitle{From conservation to crowdsourcing: A typology of
  citizen science}. In \bibinfo{booktitle}{\emph{2011 44th Hawaii international
  conference on system sciences}}. IEEE, \bibinfo{pages}{1--10}.
\newblock


\bibitem[\protect\citeauthoryear{{Wikimedia Foundation}}{{Wikimedia
  Foundation}}{2020a}]%
        {ListOfCoronaAppswikipedia_2020}
\bibfield{author}{\bibinfo{person}{{Wikimedia Foundation}}.}
  \bibinfo{year}{2020}\natexlab{a}.
\newblock \bibinfo{booktitle}{\emph{COVID-19 apps}}.
\newblock
\urldef\tempurl%
\url{https://en.wikipedia.org/wiki/COVID-19_apps}
\showURL{%
Retrieved September 14, 2020 from \tempurl}


\bibitem[\protect\citeauthoryear{{Wikimedia Foundation}}{{Wikimedia
  Foundation}}{2020b}]%
        {COVIDSymptomStudywikipedia_2020}
\bibfield{author}{\bibinfo{person}{{Wikimedia Foundation}}.}
  \bibinfo{year}{2020}\natexlab{b}.
\newblock \bibinfo{booktitle}{\emph{COVID Symptom Study}}.
\newblock
\urldef\tempurl%
\url{https://en.wikipedia.org/wiki/COVID_Symptom_Study}
\showURL{%
Retrieved September 14, 2020 from \tempurl}


\bibitem[\protect\citeauthoryear{{Wikimedia Foundation}}{{Wikimedia
  Foundation}}{2020c}]%
        {covidsafewikipedia_2020}
\bibfield{author}{\bibinfo{person}{{Wikimedia Foundation}}.}
  \bibinfo{year}{2020}\natexlab{c}.
\newblock \bibinfo{booktitle}{\emph{COVIDSafe}}.
\newblock
\urldef\tempurl%
\url{https://en.wikipedia.org/wiki/COVIDSafe}
\showURL{%
Retrieved September 14, 2020 from \tempurl}


\bibitem[\protect\citeauthoryear{{Wikimedia Foundation}}{{Wikimedia
  Foundation}}{2020d}]%
        {tracetogetherwikipedia_2020}
\bibfield{author}{\bibinfo{person}{{Wikimedia Foundation}}.}
  \bibinfo{year}{2020}\natexlab{d}.
\newblock \bibinfo{booktitle}{\emph{TraceTogether}}.
\newblock
\urldef\tempurl%
\url{https://en.wikipedia.org/wiki/TraceTogether}
\showURL{%
Retrieved September 14, 2020 from \tempurl}


\bibitem[\protect\citeauthoryear{worldometers.info}{worldometers.info}{2020}]%
        {worldometer2020}
\bibfield{author}{\bibinfo{person}{worldometers.info}.}
  \bibinfo{year}{2020}\natexlab{}.
\newblock \bibinfo{booktitle}{\emph{COVID-19 Coronavirus Pandemic}}.
\newblock
\urldef\tempurl%
\url{https://www.worldometers.info/coronavirus/}
\showURL{%
Retrieved September 14, 2020 from \tempurl}


\bibitem[\protect\citeauthoryear{{zdf.de}}{{zdf.de}}{2020a}]%
        {ZDF2020datenspendeOnline}
\bibfield{author}{\bibinfo{person}{{zdf.de}}.}
  \bibinfo{year}{2020}\natexlab{a}.
\newblock \bibinfo{booktitle}{\emph{RKI stellt App für Krankheitssymptome
  vor}}.
\newblock
\urldef\tempurl%
\url{https://www.zdf.de/nachrichten/panorama/coronavirus-rki-102.html}
\showURL{%
Retrieved December 17, 2020 from \tempurl}


\bibitem[\protect\citeauthoryear{{zdf.de}}{{zdf.de}}{2020b}]%
        {ZDF2020datenspendeTV}
\bibfield{author}{\bibinfo{person}{{zdf.de}}.}
  \bibinfo{year}{2020}\natexlab{b}.
\newblock \bibinfo{booktitle}{\emph{RKI veröffentlicht "Corona-Datenspende"
  App}}.
\newblock
\urldef\tempurl%
\url{https://www.zdf.de/nachrichten/heute-19-uhr/videos/corona-datenspende-app-100.html}
\showURL{%
Retrieved December 17, 2020 from \tempurl}


\bibitem[\protect\citeauthoryear{Zooniverse}{Zooniverse}{2020}]%
        {zooniverse}
\bibfield{author}{\bibinfo{person}{Zooniverse}.}
  \bibinfo{year}{2020}\natexlab{}.
\newblock \bibinfo{booktitle}{\emph{Zooniverse}}.
\newblock
\urldef\tempurl%
\url{https://www.zooniverse.org/projects/zookeeper/galaxy-zoo/}
\showURL{%
Retrieved September 14, 2020 from \tempurl}


\end{thebibliography}

\end{document}